\PassOptionsToPackage{unicode}{hyperref}
\PassOptionsToPackage{hyphens}{url}
\PassOptionsToPackage{dvipsnames,svgnames,x11names}{xcolor}
\documentclass[
  11pt,
]{article}
\usepackage{xcolor}
\usepackage[margin=1in]{geometry}
\usepackage{amsmath,amssymb}
\usepackage{iftex}
\ifPDFTeX
  \usepackage[T1]{fontenc}
  \usepackage[utf8]{inputenc}
  \usepackage{textcomp} 
\else 
  \usepackage{unicode-math} 
  \defaultfontfeatures{Scale=MatchLowercase}
  \defaultfontfeatures[\rmfamily]{Ligatures=TeX,Scale=1}
\fi
\usepackage{lmodern}
\ifPDFTeX\else
\fi
\IfFileExists{upquote.sty}{\usepackage{upquote}}{}
\IfFileExists{microtype.sty}{
  \usepackage[]{microtype}
  \UseMicrotypeSet[protrusion]{basicmath} 
}{}
\makeatletter
\@ifundefined{KOMAClassName}{
  \IfFileExists{parskip.sty}{%
    \usepackage{parskip}
  }{
    \setlength{\parindent}{0pt}
    \setlength{\parskip}{6pt plus 2pt minus 1pt}}
}{
  \KOMAoptions{parskip=half}}
\makeatother
\usepackage{longtable,booktabs,array}
\usepackage{calc} 
\usepackage{etoolbox}
\makeatletter
\patchcmd\longtable{\par}{\if@noskipsec\mbox{}\fi\par}{}{}
\makeatother
\IfFileExists{footnotehyper.sty}{\usepackage{footnotehyper}}{\usepackage{footnote}}
\makesavenoteenv{longtable}
\usepackage{graphicx}
\makeatletter
\newsavebox\pandoc@box
\newcommand*\pandocbounded[1]{
  \sbox\pandoc@box{#1}%
  \Gscale@div\@tempa{\textheight}{\dimexpr\ht\pandoc@box+\dp\pandoc@box\relax}%
  \Gscale@div\@tempb{\linewidth}{\wd\pandoc@box}%
  \ifdim\@tempb\p@<\@tempa\p@\let\@tempa\@tempb\fi
  \ifdim\@tempa\p@<\p@\scalebox{\@tempa}{\usebox\pandoc@box}%
  \else\usebox{\pandoc@box}%
  \fi%
}
\def\fps@figure{htbp}
\makeatother
\providecommand{\tightlist}{%
  \setlength{\itemsep}{0pt}\setlength{\parskip}{0pt}}
\usepackage{seqsplit}

\usepackage{newunicodechar}

\newunicodechar{α}{\ensuremath{\alpha}}
\newunicodechar{β}{\ensuremath{\beta}}
\newunicodechar{γ}{\ensuremath{\gamma}}
\newunicodechar{δ}{\ensuremath{\delta}}
\newunicodechar{ε}{\ensuremath{\varepsilon}}
\newunicodechar{η}{\ensuremath{\eta}}
\newunicodechar{θ}{\ensuremath{\theta}}
\newunicodechar{κ}{\ensuremath{\kappa}}
\newunicodechar{λ}{\ensuremath{\lambda}}
\newunicodechar{μ}{\ensuremath{\mu}}
\newunicodechar{ν}{\ensuremath{\nu}}
\newunicodechar{π}{\ensuremath{\pi}}
\newunicodechar{ρ}{\ensuremath{\rho}}
\newunicodechar{σ}{\ensuremath{\sigma}}
\newunicodechar{τ}{\ensuremath{\tau}}
\newunicodechar{φ}{\ensuremath{\varphi}}
\newunicodechar{χ}{\ensuremath{\chi}}
\newunicodechar{ψ}{\ensuremath{\psi}}
\newunicodechar{ω}{\ensuremath{\omega}}

\newunicodechar{Δ}{\ensuremath{\Delta}}
\newunicodechar{Σ}{\ensuremath{\Sigma}}
\newunicodechar{Π}{\ensuremath{\Pi}}
\newunicodechar{Ω}{\ensuremath{\Omega}}

\newunicodechar{∈}{\ensuremath{\in}}
\newunicodechar{∉}{\ensuremath{\notin}}
\newunicodechar{≈}{\ensuremath{\approx}}
\newunicodechar{≤}{\ensuremath{\leq}}
\newunicodechar{≥}{\ensuremath{\geq}}
\newunicodechar{≠}{\ensuremath{\neq}}
\newunicodechar{∼}{\ensuremath{\sim}}
\newunicodechar{∞}{\ensuremath{\infty}}
\newunicodechar{×}{\ensuremath{\times}}
\newunicodechar{·}{\ensuremath{\cdot}}
\newunicodechar{±}{\ensuremath{\pm}}
\newunicodechar{−}{\ensuremath{-}}
\newunicodechar{∝}{\ensuremath{\propto}}
\newunicodechar{∑}{\ensuremath{\sum}}
\newunicodechar{∫}{\ensuremath{\int}}
\newunicodechar{∂}{\ensuremath{\partial}}

\newunicodechar{→}{\ensuremath{\rightarrow}}
\newunicodechar{←}{\ensuremath{\leftarrow}}
\newunicodechar{↔}{\ensuremath{\leftrightarrow}}
\newunicodechar{⇒}{\ensuremath{\Rightarrow}}
\newunicodechar{⇐}{\ensuremath{\Leftarrow}}

\newunicodechar{★}{\ensuremath{\star}}

\newunicodechar{ä}{\"a}
\newunicodechar{ü}{\"u}
\newunicodechar{ö}{\"o}
\newunicodechar{é}{\'e}
\newunicodechar{è}{\`e}

\usepackage{bookmark}
\IfFileExists{xurl.sty}{\usepackage{xurl}}{} 
\providecommand{\xmpquote}[1]{#1}
\hypersetup{
  pdftitle={Benchmarking Recursive-Collapse Warning Claims Under Matched False-Positive Control},
  pdfauthor={David Mullett},
  pdfsubject={Matched false-positive benchmarking for recursive-collapse warning claims},
  pdfkeywords={\xmpquote{recursive collapse, matched false-positive, early warning, benchmark framework, pre-collapse, falsifiable}},
  colorlinks=true,
  linkcolor={blue},
  filecolor={Maroon},
  citecolor={blue},
  urlcolor={blue},
  pdfcreator={LaTeX via pandoc}}

\author{}
\date{}

\begin{document}

\section{Benchmarking Recursive-Collapse Warning Claims Under Matched
False-Positive
Control}\label{benchmarking-recursive-collapse-warning-claims-under-matched-false-positive-control}

\textbf{David Mullett}

Independent Researcher

ORCID: 0009-0004-2543-1664

\textbf{Corresponding author:} David Mullett ·
\href{mailto:d@loopzero.org}{\nolinkurl{d@loopzero.org}}

\begin{center}\rule{0.5\linewidth}{0.5pt}\end{center}

\subsection{One-Sentence Summary}\label{one-sentence-summary}

A matched-FP benchmark for recursive collapse: signature directionally
aligned across domains; no detector accepted.

\begin{center}\rule{0.5\linewidth}{0.5pt}\end{center}

\subsection{Abstract}\label{abstract}

Recursive systems can enter collapse-like regimes --- self-reinforcing
amplification, persistent recursion, and narrowing diversity that mask
accelerating internal degradation --- before overt failure becomes
visible. We introduce Loopzero, a claim-bounded benchmark framework for
testing whether recursive failures follow a directional telemetry
pattern: rising gain (G), recursive persistence (p), and declining
diversity (δ). The claim boundary is specified in Lean; the Lean
artifact does not verify real telemetry, benchmark validity, or detector
performance.

We evaluate the bridge on two frozen public-artifact benchmarks: a
segmented public-markets benchmark (Volmageddon 2018, COVID MWCB 2020)
and a MovieLens-25M offline deterministic recommender replay. Detectors
are evaluated under a locked equal-false-positive contract (FP ∈
{[}0.03, 0.07{]}, pre-registered) so all configurations face the same
alert budget. Neither tested standard comparators nor Loopzero's
pre-registered quantile detector achieved an accepted operating point.
Directional witness alignment held on both canonical benchmarks, with
adjacent-horizon and row-level limitations disclosed. Digitized
Shumailov et al.~(2024) LLM training-loop trajectories are directionally
consistent with the pattern; matched-FP evaluation in that domain is
deferred.

The contribution is a reproducible, falsifiable benchmark framework for
evaluating recursive-collapse warning claims under an explicit
alert-budget contract --- non-acceptance reported as a first-class
scientific outcome.

\begin{center}\rule{0.5\linewidth}{0.5pt}\end{center}

\pandocbounded{\includegraphics[keepaspectratio]{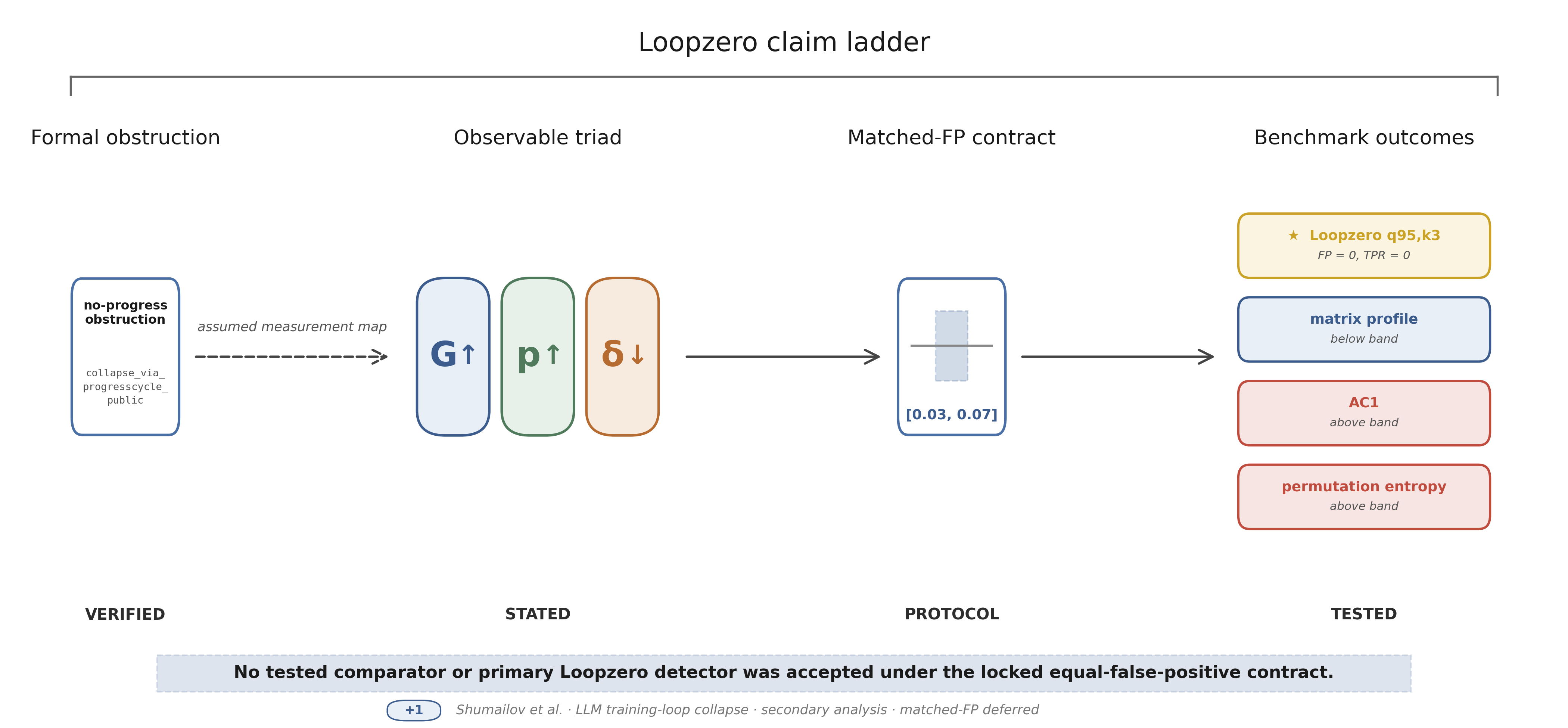}}

\textbf{Figure 1. Loopzero claim ladder: from Lean-checked obstruction
to empirical benchmark claims.} Across the canonical public-markets and
MovieLens-25M recommender benchmarks, no tested comparator configuration
achieved acceptance under the locked equal-false-positive band {[}0.03,
0.07{]}: matrix profile (FP = 0.014, under-band) and AC1 (FP = 0.132,
over-band) are illustrative cases shown; permutation entropy (FP =
0.368, far above band) completes the calibration outcome. The figure
decomposes the underlying claim into four epistemic layers: (i) an
elementary no-progress obstruction verified in Lean
(\texttt{\seqsplit{collapse\_via\_progresscycle\_public}}); (ii) an assumed
measurement map to observable G/p/δ telemetry, stated as a conditional
bridge; (iii) the matched-false-positive protocol; and (iv) empirical
testing on frozen public-artifact benchmarks. Comparator markers are
color-coded by failure mode (blue = under-band; orange = over-band).
Sample sizes: canonical segmented markets n\_control=38 + n\_event=16;
canonical recommender at h=50 n\_control=4,755 + n\_event=35,584 (40,339
user clusters total). The Shumailov et al.~model-collapse analysis is
included as a secondary directional-consistency check only; matched-FP
evaluation in that domain is deferred.

\subsection{Introduction}\label{introduction}

\subsubsection{A formally specified obstruction for recursive
collapse}\label{a-formally-specified-obstruction-for-recursive-collapse}

Many systems of scientific and practical interest are recursive: present
state shapes future state, and locally valid updates can reinforce their
own continuation. Collapse, catastrophic shift, and resilience loss are
well-established phenomena in complex systems (Scheffer et al., 2001;
Holling, 1973). In such recursive systems, collapse need not begin as
overt breakdown. Related recursive-collapse behavior has recently been
demonstrated in generative-model training loops, where repeated training
on recursively generated data degrades model behavior (Shumailov et al.,
2024). Building on the established early-warning and critical slowing
down literature, which has developed variance growth, autocorrelation
shifts, and threshold excursions as observables for critical-transition
detection (Scheffer et al., 2009; Scheffer et al., 2012; Carpenter et
al., 2011; Dakos et al., 2012), this paper begins from a complementary
starting point: a formally specified structural obstruction
characterizing one specific failure mode within the broader
recursive-collapse landscape. The empirical evaluation focuses on
digital recursive systems --- segmented public-markets telemetry and an
offline deterministic recommender --- with the Shumailov training-loop
setting included as a secondary consistency check; the formal
obstruction itself is intended to apply more broadly.

We formalize this as a no-progress obstruction for recursive feedback
systems, verified in Lean (de Moura \& Ullrich, 2021; The mathlib
Community, 2020). In that formal setting, one strictly worsening update
coupled to monotone tethers can generate a cycle in which local
continuation remains possible while meaningful progress is lost. In
plain terms, a no-progress obstruction is a condition under which a
system keeps taking valid local steps yet never actually improves. A
monotone tether is a quantity constrained to move in one direction only
--- able to hold or worsen, but not recover. The empirical question is
therefore not whether one can engineer a useful detector from arbitrary
signals. It is whether this no-progress obstruction admits a conditional
bridge, under stated assumptions, to a measurable pre-collapse footprint
in observed recursive systems.

The formal obstruction is intentionally elementary; the Lean development
is not presented as a substantive contribution in formal methods. Its
purpose is to specify the claim boundary in machine-checkable form, so
that the formal/empirical divide is explicit and inspectable. The Lean
artifact contains three components: an abstract no-progress obstruction
over preorders, a measurement-map bridge, and a schematic G,p,δ-style
telemetry specialization. A preorder here is simply a ranking of system
states by `at least as good as' that allows ties and pairs that need not
be directly comparable. The measurement map is the assumed link from
observable telemetry to that ranking, which the paper assumes rather
than proves. Lean verifies the structural ordered-measurement
obstruction and the schematic telemetry specialization; it does not
verify that real benchmark telemetry supplies the required measurement
map. The empirical claims in this paper depend on the bridge protocol
described below, not on the formal artifact.

\subsubsection{Conditional telemetry
bridge}\label{conditional-telemetry-bridge}

The formal result does not itself specify a unique empirical measurement
scheme for collapse. Cascading failure in interdependent systems
provides structural motivation for treating recursive no-progress as a
distinct failure mode (Buldyrev et al., 2010). The formal result
identifies a class of recursive no-progress regimes in which local
continuation remains possible while meaningful forward improvement
becomes inaccessible. We therefore treat the bridge as conditional
rather than automatic: to relate the obstruction to data, we consider
recursive systems in which future updates depend partly on recent
internal outputs, partly on external input, and partly on the diversity
of accessible next states. Under this interpretation, collapse need not
first appear as abrupt breakdown. It may instead emerge as an observable
pre-collapse regime in which perturbations are increasingly amplified,
recent internally generated states increasingly determine subsequent
states, and the accessible state space contracts.

The empirical claim that real benchmark telemetry supplies such a
measurement map remains external to Lean. It is tested by the benchmark
protocol: the observed G,p,δ witnesses must satisfy the prespecified
bridge criterion before externally defined collapse events under the
locked false-positive contract, while comparator families are evaluated
under the same alert budget.

The obstruction therefore motivates three testable observable tendencies
before externally defined collapse events. First, rising gain G,
indicating amplification rather than damping of perturbation. Second,
rising recursive persistence p (operationalized in the bridge criterion
as a non-relaxation gate), indicating increasing persistence of
internally generated state. Third, declining diversity δ, indicating
contraction of the effective range of system trajectories (Walker et
al., 2004; Gao et al., 2016). We use G, p, and δ as empirical proxies
for amplification, recursive persistence, and state-space contraction.
This bridge is explicitly falsifiable. It would be weakened if
benchmark-defined collapse repeatedly occurred without prior elevation
of G and p and reduction of δ, or if matched-false-positive comparator
configurations repeatedly recovered accepted operating points without
this triad.

\emph{Scope summary.} (i) Lean verifies the abstract no-progress
obstruction over preorders, the bridge lemma
\texttt{\seqsplit{collapse\_via\_progresscycle\_public}}, and a schematic telemetry
specialization. (ii) The conditional bridge assumes that real recursive
systems admit a measurement function μ mapping observable telemetry into
a preorder, under which the witness triad's directional structure
(rising G, rising p, declining δ) corresponds to the formal no-progress
condition. (iii) The benchmarks test whether the witness pattern in fact
precedes externally defined event labels under the locked
equal-false-positive contract; this is an empirical question, not a
verified one.

\subsubsection{Telemetry witnesses and matched-false-positive
benchmarking}\label{telemetry-witnesses-and-matched-false-positive-benchmarking}

The bridge criterion is operationalized through a triplet of witnesses.
G estimates short-horizon amplification of perturbation, p estimates
short-horizon recursive persistence of active state, and δ estimates the
effective breadth of observed system behavior across channels, modes, or
trajectories. The predicate is therefore designed to test the empirical
signature implied by the formal mechanism, not to restate the formal
obstruction in observational language.

In the current implementation, the three witnesses are not equally
directional in their operational role. G is evaluated as an
elevated-gain condition relative to recent background, and δ is
evaluated through a non-increase condition over the relevant lookback
window. The p witness is used most conservatively: it functions as a
non-relaxation gate consistent with monotone-tether semantics rather
than as a standalone directional effect-size claim. In the
public-markets adapter, p is constructed from a trailing recurrence
statistic over stress events defined using a fixed z-threshold of 1.5;
this threshold is part of the current adapter definition and should be
interpreted as such.

All empirical evaluation is conducted under a matched false-positive
contract. In heterogeneous systems, comparator families can appear
competitive simply by spending false alarms differently (Boettiger \&
Hastings, 2012a; Jäger \& Füllsack, 2019). We therefore compare
detectors only at the same alert budget, defined by a locked
equal-false-positive criterion rather than by arbitrary family-specific
thresholds (Boettiger \& Hastings, 2012b; Hanley \& McNeil, 1982). A
configuration is accepted only if its control-unit false-positive rate
lies within the prespecified interval FP ∈ {[}0.03, 0.07{]}.

The acceptance band was chosen prior to evaluation to reflect the
operational alert burden plausible for an operator monitoring a
comparator unit population over a comparable observation period. A mean
false-positive rate of 0.05 corresponds to approximately one false alert
per 20 control-period units; the half-width of ±0.02 reflects acceptable
sampling variation under finite-control evaluation. Tighter bands such
as {[}0.01, 0.05{]} would be operationally desirable but require larger
control-unit populations than the canonical markets benchmark provides;
wider bands such as {[}0.05, 0.10{]} would inflate the false-alarm
budget beyond what is operationally tolerable in a deployed warning
context. The band was locked at this specification before any comparator
evaluation. We note that on the canonical markets benchmark with n=38
controls, the grid step 1/38 = 0.026316 admits only 2/38 control alarms
within the band (FP = 0.052632), making the band effectively a
single-grid-point target on that benchmark; the recommender benchmark
with n=4,755 controls (grid step 1/4755 = 0.0002103) is therefore the
empirically dispositive case, and the markets non-acceptance result
should be read as corroborating second-domain evidence rather than as an
independent statistical claim of comparator inadequacy.

Under this contract, the comparator-non-recovery claim is directly
falsifiable: it fails if any comparator configuration admits an accepted
operating point on the canonical benchmark under the same criterion. An
operating point is one setting at which a detector runs, fixing its
false-positive and true-positive rates. It counts as accepted only if
its false-positive rate falls inside the locked band {[}0.03, 0.07{]}.

\subsection{Results}\label{results}

\subsubsection{Cross-domain evidence for a pre-collapse
signature}\label{cross-domain-evidence-for-a-pre-collapse-signature}

We therefore test whether the witness triad (G, p, δ) appears before
externally defined collapse events under a locked equal-false-positive
benchmark across heterogeneous recursive systems. Across retained domain
families, the witness triad aligns directionally ahead of externally
defined breakdown or intervention points and does so under the same
alert-budget rule. These retained families were chosen not because they
are easy cases, but because they admit externally defined event times,
nontrivial controls, and fair comparator evaluation under a common
false-positive contract.

Directional witness summaries were fully aligned on the canonical
recommender benchmark. On the canonical public-markets benchmark, the
same directional pattern was recovered when markets were summarized over
the last 30 minutes of each exact canonical unit using G, p, and
diversity-change within the late window. In a wider 60-minute markets
summary, G and p remained directionally aligned whereas diversity-change
weakened. We therefore interpret the markets bridge as localized and
late-window strongest rather than uniformly invariant across wider
summary windows.

Per-witness effect-size summaries extend this directional analysis to
magnitude and uncertainty. We compute Cohen's d, Glass's d, and rank AUC
for each witness (G, p, δ) at each benchmark/horizon, with cluster-aware
bootstrap 95\% BCa confidence intervals over 10,000 iterations (markets:
segment-level resampling, n=38 controls + 16 events; recommender:
user-level resampling, n = 40,339 user clusters). At the canonical
recommender horizon h=50, all three witnesses point in the predicted
direction with intervals clear of the null reference (G Cohen's d =
+0.10 {[}+0.08, +0.12{]}; p Cohen's d = +0.08 {[}+0.05, +0.11{]}; δ
Cohen's d = −0.17 {[}−0.21, −0.13{]}). At adjacent horizons the pattern
degrades asymmetrically: at h=40 the G witness flips to a
wrong-direction effect (d = −0.21) while p and δ hold in the predicted
direction; at h=60 G strengthens further (d = +0.33) while p and δ
collapse to null intervals. The pre-registered canonical horizon is
therefore the only configuration where all three witnesses align in the
predicted direction with intervals clear of null. On the markets
benchmark, per-row effect sizes are uniformly null across all three
witnesses (intervals span the null reference); the markets directional
pattern reported above is recovered at the unit-level aggregation grain
used in the late-window analysis (mean per segment, then compare across
n=38+16 segments), not at the per-row grain on which these effect sizes
are computed. A full effect-size table with both BCa and percentile CIs
is reported in Supplementary Table S2; Figure 6 provides a forest-plot
visualization.

The broader significance is not that a new warning statistic works in
one domain. It is that a no-progress obstruction appears to support a
domain-resolved empirical signature across systems that differ strongly
in substrate, timescale, and external semantics. The evidence therefore
supports an explicitly bounded empirical program rather than a
single-domain detector construction.

A domain-resolved witness-direction summary is provided in the
Supplementary Materials. The combined comparator calibration summary
across both flagship benchmarks is reported in Table 1.

\textbf{Table 1. Matched false-positive evaluation on frozen
public-artifact benchmarks under the locked equal-false-positive
contract (FP ∈ {[}0.03, 0.07{]}).} Panels: markets (n=38 controls + 16
events on volmageddon\_covid\_public\_v2) and canonical recommender
(n=4,755 controls + 35,584 events at horizon h=50 on
movielens25m\_recursive\_frontier\_public\_v1; 40,339 user clusters
total). Standard comparator families and Loopzero's pre-registered
quantile detector were evaluated under the same locked
equal-false-positive contract. No tested comparator configuration and no
operating point of Loopzero's pre-registered quantile detector achieved
an accepted operating point on either benchmark. The directional G/p/δ
bridge summaries are therefore reported separately from binary detector
acceptance.

Loopzero's pre-registered quantile detector (G \textgreater{} 95th
percentile, p \textgreater{} 95th percentile, δ \textless{} 5th
percentile of control-unit reference rows; k = 3 consecutive windows) is
evaluated at its pre-registered primary operating point on both
canonical benchmarks. On the canonical markets benchmark (1,120
reference rows from 38 control units), the detector alarms 0 of 38
control units (FP = 0.000) and 0 of 16 event units (event alarm rate =
0.000). On the canonical recommender benchmark (47,550 reference rows
from 4,755 control units), the detector alarms 0 of 4,755 control units
(FP = 0.000) and 0 of 35,584 event units (event alarm rate = 0.000). The
pre-registered operating point therefore falls under-band on both
benchmarks. No tested operating point in either benchmark's locked
equal-FP band {[}0.03, 0.07{]} achieves event recovery; full 9-cell q ×
k sensitivity grids per benchmark are reported in
\texttt{\seqsplit{results/rendered/bridge/a1\_loopzero\_operating\_points.md}}, and
the pre-registered specification is documented in
\texttt{\seqsplit{analysis/14\_a1\_prereg.md}}. Under matched-FP discipline,
Loopzero's pre-registered detector and the tested standard comparator
families therefore both report non-acceptance on the canonical
benchmarks. This consistency is itself a coherent outcome of the
matched-FP framework, which evaluates detectors on a level playing field
rather than at family-specific thresholds; the manuscript's central
comparator-non-recovery claim and the bridge-criterion directional
analysis (witnesses G, p, δ separating event from control rows in the
prespecified directions on the canonical recommender benchmark and on
the canonical late-window markets summary) are unaffected by this
detector-specific non-acceptance result.

{\def\LTcaptype{none} 
\begin{longtable}[]{@{}
  >{\raggedright\arraybackslash}p{(\linewidth - 12\tabcolsep) * \real{0.1429}}
  >{\raggedright\arraybackslash}p{(\linewidth - 12\tabcolsep) * \real{0.1429}}
  >{\raggedright\arraybackslash}p{(\linewidth - 12\tabcolsep) * \real{0.1429}}
  >{\raggedright\arraybackslash}p{(\linewidth - 12\tabcolsep) * \real{0.1429}}
  >{\raggedright\arraybackslash}p{(\linewidth - 12\tabcolsep) * \real{0.1429}}
  >{\raggedright\arraybackslash}p{(\linewidth - 12\tabcolsep) * \real{0.1429}}
  >{\raggedright\arraybackslash}p{(\linewidth - 12\tabcolsep) * \real{0.1429}}@{}}
\toprule\noalign{}
\begin{minipage}[b]{\linewidth}\raggedright
Benchmark
\end{minipage} & \begin{minipage}[b]{\linewidth}\raggedright
Comparator (family, role)
\end{minipage} & \begin{minipage}[b]{\linewidth}\raggedright
Configuration
\end{minipage} & \begin{minipage}[b]{\linewidth}\raggedright
Control FP
\end{minipage} & \begin{minipage}[b]{\linewidth}\raggedright
Event alarm rate
\end{minipage} & \begin{minipage}[b]{\linewidth}\raggedright
Distance from band
\end{minipage} & \begin{minipage}[b]{\linewidth}\raggedright
Status
\end{minipage} \\
\midrule\noalign{}
\endhead
\bottomrule\noalign{}
\endlastfoot
Markets (n=38 controls, 16 events) & AC1 (fast --- nearest nontrivial) &
\texttt{\seqsplit{ac1\_ews\_\_632f23b2}} & 0.1316 (5/38) & 0.0625 (1/16) & +0.0616
above & Above band \\
Markets & Permutation entropy (slow --- best nontrivial) &
\texttt{\seqsplit{permutation\_entropy\_\_259c1b96}} & 0.3684 (14/38) & 0.2500
(4/16) & +0.2984 above & Above band \\
Markets & Matrix profile / permutation entropy (slow --- numerically
nearest) & trivial-silent ties & 0.0000 (0/38) & 0.0000 (0/16) & --- &
Trivial-silent \\
Recommender (n=4,755 controls, 35,584 events; 105 configurations tested)
& Matrix profile (slow --- overall nearest) &
\texttt{\seqsplit{matrix\_profile\_\_44b81bd6}} & 0.01367 & 0.1107 & −0.0163 below
& Below band \\
Recommender & Variance EWS (fast --- nearest) & trivial-silent & 0.0000
& --- & --- & Trivial-silent \\
\end{longtable}
}

\emph{No tested comparator configuration achieved an accepted operating
point on either benchmark. Comparator calibration under the matched
false-positive contract is not equivalent to universal detector ranking;
comparators and Loopzero's pre-registered detector may admit competitive
operating points under different evaluation frameworks. The contract
evaluates whether the prespecified canonical band can be reached, not
which detector is universally superior.}

\subsubsection{Canonical public recommender
benchmark}\label{canonical-public-recommender-benchmark}

A second flagship benchmark was constructed from MovieLens-25M user
trajectories to test whether the bridge criterion survives outside
markets. Recommender systems are known to narrow accessible diversity
through feedback-driven filtering (Fleder \& Hosanagar, 2009). User
ratings were sorted chronologically, reduced to one episode per user,
and replayed under a deterministic item-item collaborative-filtering
engine with a warm-start prefix and a held-out positive frontier (Chaney
et al., 2018; Jiang et al., 2019; Mansoury et al., 2020). On the
canonical 50-step benchmark, 40,339 user-level units satisfied inclusion
criteria, including 35,584 event units and 4,755 control units. The
control-unit false-positive grid step was 1/4755 = 0.0002103, so the
prespecified acceptance band {[}0.03, 0.07{]} was reachable by
construction.

On this canonical recommender benchmark, the witness pattern satisfied
the prespecified bridge criterion in the pre-collapse window:
pre-collapse event units showed higher amplification G, higher recursive
persistence p, and lower diversity δ than reference controls. User-level
bootstrap summaries were directionally consistent with this pattern and
were used descriptively rather than as a significance test. Under the
same locked equal-false-positive contract, no tested fast or slow
comparator configuration admitted an accepted operating point. Across
105 tested configurations, the overall nearest comparator was matrix
profile (matrix\_profile\_\_44b81bd6), which remained below the lower
edge of the accepted band with control FP = 0.0136698, band distance =
0.0163302, and event alarm rate = 0.110668. Among fast families, the
numerically nearest operating point was a trivial-silent variance EWS
configuration at FP = 0.0, whereas the nearest nontrivial fast-family
configurations overfired controls substantially.

\pandocbounded{\includegraphics[keepaspectratio]{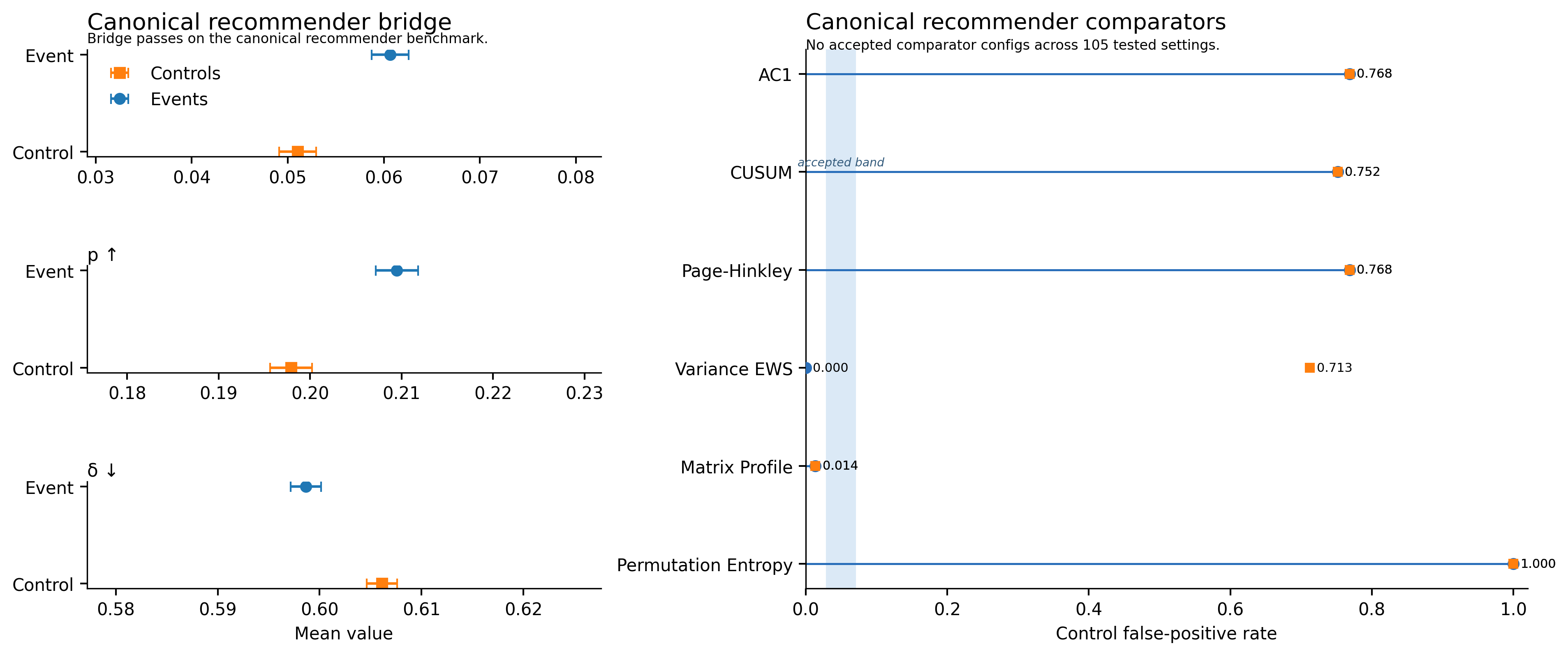}}

\textbf{Figure 2. Canonical recommender benchmark: bridge summary and
comparator calibration.} On the canonical 50-step MovieLens-25M
recursive frontier benchmark (n=4,755 control units + 35,584 event units
= 40,339 user clusters total at horizon h=50), directional witness
summaries are fully aligned: event units show higher gain G, higher
recursive persistence p, and lower diversity δ than controls. Under the
same locked equal-false-positive contract, no tested comparator
configuration is accepted across 105 tested settings; the overall
nearest comparator is matrix profile at FP = 0.014, below the lower edge
of the accepted band. Directional witness alignment in this canonical
h=50 panel is small in magnitude (G d ≈ +0.10, p d ≈ +0.08, δ d ≈
−0.17); the bridge claim is a directional one and is not equivalent to a
claim of large discriminative effect size. Effect-size estimates and BCa
intervals are reported in Figure 6 and Supplementary Table S2. Results
are based on offline deterministic replay rather than deployed-system
feedback.

\subsubsection{Recommender robustness under adjacent horizon
sensitivity}\label{recommender-robustness-under-adjacent-horizon-sensitivity}

To test whether the recommender result depended on the canonical episode
horizon, the benchmark was rebuilt at adjacent horizons of 40 and 60
recursive update steps while holding benchmark construction and
comparator rules fixed. At both adjacent horizons, no comparator
configuration recovered an accepted operating point under the locked
equal-false-positive band. The overall nearest comparator remained
matrix profile in all three horizon settings and remained outside the
accepted interval. The bridge degrades asymmetrically across adjacent
horizons. At h=40, witnesses p and δ remain aligned in the predicted
direction with bootstrap CIs clear of null, but G's sign reverses
(events show lower gain than controls, opposite to the prediction). At
h=60, G's alignment strengthens, but bootstrap CIs for p and δ collapse
to include null. The pre-registered canonical horizon h=50 is therefore
the only tested configuration where all three witnesses simultaneously
align in the predicted direction with bootstrap CIs clear of null
(Supplementary Table S2). This asymmetric horizon dependence is
consistent with witness-specific timescales --- G is sensitive to
cumulative amplification (favoring longer horizons) while p and δ
measure within-trajectory recursive structure that becomes harder to
distinguish from control trajectories at longer horizons --- and
reinforces the load-bearing role of pre-registration for the bridge
claim. Loopzero's pre-registered quantile detector was also evaluated at
its primary operating point (q=95, k=3) on both adjacent-horizon
packets. On the recommender benchmark rebuilt at h=40 (65,620 reference
rows from 6,562 control units), the detector alarms 0 of 6,562 control
units (FP = 0.000) and 0 of 33,777 event units (event alarm rate =
0.000). On the recommender benchmark rebuilt at h=60 (39,960 reference
rows from 3,996 control units), the detector alarms 0 of 3,996 control
units (FP = 0.000) and 0 of 36,343 event units (event alarm rate =
0.000). The detector's primary-cell non-acceptance pattern observed at
h=50 therefore extends across all three tested horizons; full 9-cell q ×
k sensitivity grids at h=40 and h=60 are reported in
\texttt{\seqsplit{results/rendered/bridge/a1\_loopzero\_operating\_points\_h40\_h60.md}}.
The recommender branch therefore supports a corroborating second
flagship benchmark on the comparator claim, with a bounded robustness
qualification on the bridge layer under adjacent horizon shortening
rather than an unqualified invariance claim.

\pandocbounded{\includegraphics[keepaspectratio]{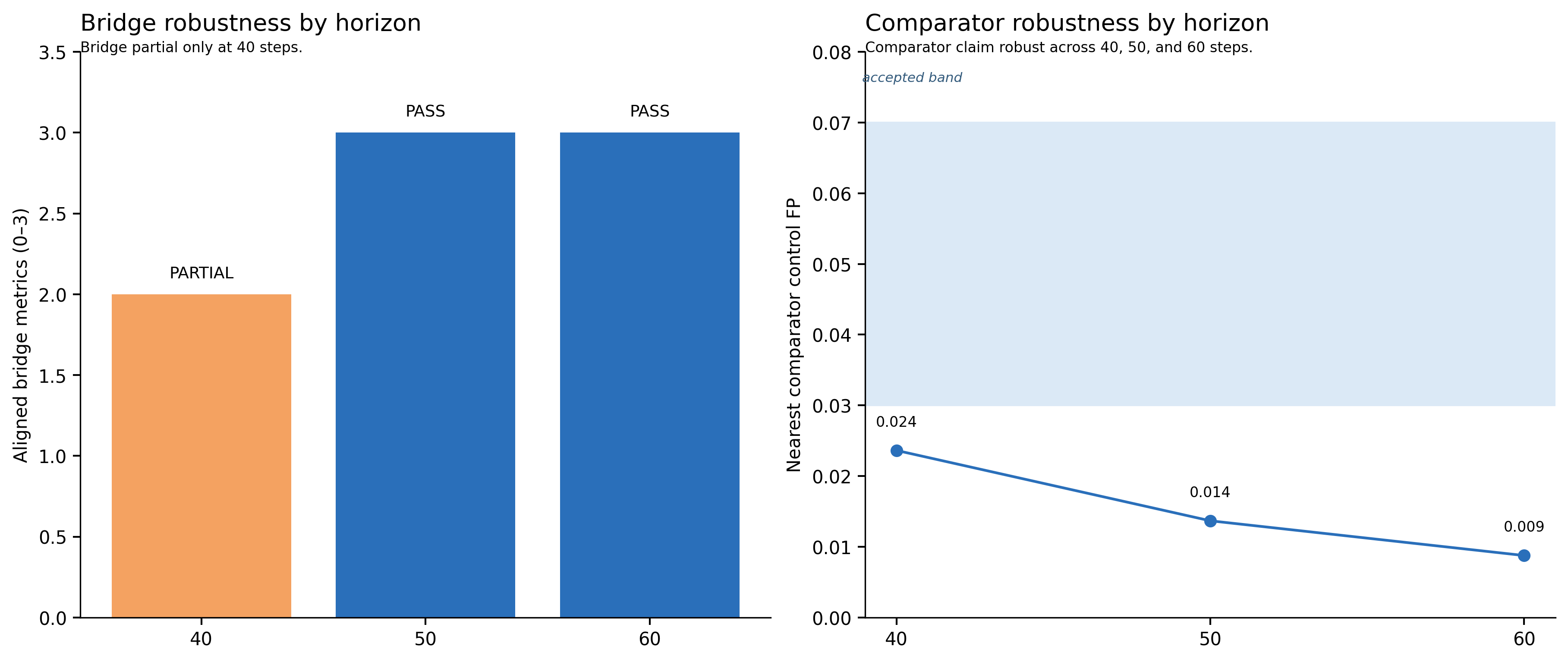}}

\textbf{Figure 3. Recommender robustness by adjacent horizon
sensitivity.} Evaluation on the MovieLens-25M recursive frontier
benchmark at three horizons (40,339 user clusters per panel;
event/control splits vary by horizon --- h=40: 33,777 events + 6,562
controls; h=50 canonical: 35,584 events + 4,755 controls; h=60: 36,343
events + 3,996 controls). The bridge criterion is satisfied at 50 and 60
steps and only partially satisfied at 40 steps, indicating bounded
horizon sensitivity rather than failure of the empirical program. The
comparator claim is robust across all three horizons: no tested
comparator configuration is accepted at 40, 50, or 60 steps.
PASS/PARTIAL refers exclusively to the prespecified direction-count
bridge rule (event-unit G higher than control, p non-relaxation
maintained, event-unit δ lower than control); it does not require
effect-size intervals to exclude null. Per-horizon effect-size
magnitudes and BCa intervals are reported in Figure 6 and Supplementary
Table S2: h=40 shows G in the wrong direction (d = −0.21) with p and δ
still aligned; h=60 shows G strengthening (d = +0.33) while p and δ
effect-size intervals collapse to null. The pre-registered canonical
horizon h=50 is the only configuration where all three witnesses align
in the predicted direction with intervals clear of null. Results are
based on offline deterministic replay rather than deployed-system
feedback.

\subsubsection{Public market event
family}\label{public-market-event-family}

Public markets provide a stringent test because they combine exogenous
shocks, endogenous amplification (Brunnermeier \& Pedersen, 2009),
volatility-linked stress indicators such as the VIX (Whaley, 2009; ESRB
Advisory Scientific Committee, 2019), heterogeneous liquidity structure,
and a substantial comparator literature. We therefore assembled a
reproducible public market event family centered on the February 2018
Volmageddon dislocation (Augustin et al., 2021) and the March 2020 COVID
market-wide circuit-breaker cluster (NYSE Market-Wide Circuit Breaker
Working Group, 2021), together with hard negative controls. The purpose
was not to present markets as a universal separator domain, but to ask
whether a formally specified recursive warning criterion survives
externally auditable benchmark construction in a domain with strong
prior comparators and well-known confounds --- macro-driven volatility,
microstructure effects, calendar regularities, and news-shock dynamics
that can produce volatility excursions without any underlying recursive
feedback mechanism.

In this branch, the critical distinction is between narrative motivation
and frozen benchmark evaluation. Narrative examples such as the 2010
Flash Crash (CFTC and SEC Staff, 2010) motivate the broader problem of
recursive dislocation, but the quantitative claims reported here are
anchored to the frozen canonical markets benchmark and its comparator
contract. That benchmark is the relevant object for comparator
availability, equal-false-positive reachability, and robustness.

\subsubsection{Comparator calibration on the canonical markets
benchmark}\label{comparator-calibration-on-the-canonical-markets-benchmark}

To test whether the markets result depended on a narrow comparator set,
we evaluated a broader comparator suite on the canonical segmented
markets benchmark, volmageddon\_covid\_public\_v2, constructed from a
single canonical source configuration and a rule in which one underlying
market segment constituted one comparator unit. This yielded 38 control
units and 16 event units, so the prespecified equal-false-positive
acceptance band, FP ∈ {[}0.03, 0.07{]}, was reachable on this benchmark
with grid step 1/38 = 0.026316. Fast comparator families comprised
variance EWS and lag-1 autocorrelation (AC1) (Dakos et al., 2012), CUSUM
(Page, 1954), and Page-Hinkley (Page, 1954); slow families comprised
matrix profile (Yeh et al., 2016) and permutation entropy (Bandt \&
Pompe, 2002). All families were evaluated under the same locked
equal-false-positive contract with predeclared parameter grids, followed
by full-grid expansion of the slow families on the frozen canonical
benchmark. Two procedural filters apply during calibration: a
\texttt{trivial\_silent} flag identifies comparator configurations that
emit zero alarms across the entire benchmark and therefore satisfy any
false-positive band by vacuous compliance --- these configurations are
excluded from accepted-operating-point determination; an
\texttt{\seqsplit{unreachable\_fp\_band}} filter excludes configurations whose
minimum-length pre-conditions reduce n\_control\_units below the
discrete-grid floor required to reach the {[}0.03, 0.07{]} band,
evaluated before calibration.

No tested fast or slow comparator configuration achieved the acceptance
band on the canonical markets benchmark. The nearest nontrivial
comparator remained AC1: configuration ac1\_ews\_\_632f23b2 alarmed on
5/38 control units (FP = 0.131579) and 1/16 event units, leaving it
0.061579 above the admissible band. Full-grid slow-family evaluation did
not rescue the comparator branch. The numerically nearest slow
configurations were tied across matrix profile and permutation entropy,
but both were trivial-silent, producing 0/38 control alarms and 0/16
event alarms. Among slow families, the best nontrivial configuration
came from permutation entropy (permutation\_entropy\_\_259c1b96), which
alarmed on 14/38 control units (FP = 0.368421) and 4/16 event units,
remaining 0.298421 above band. Thus, under a reachable and locked
equal-false-positive criterion, no tested comparator configuration
admitted an acceptable operating point on the canonical markets
benchmark.

\pandocbounded{\includegraphics[keepaspectratio]{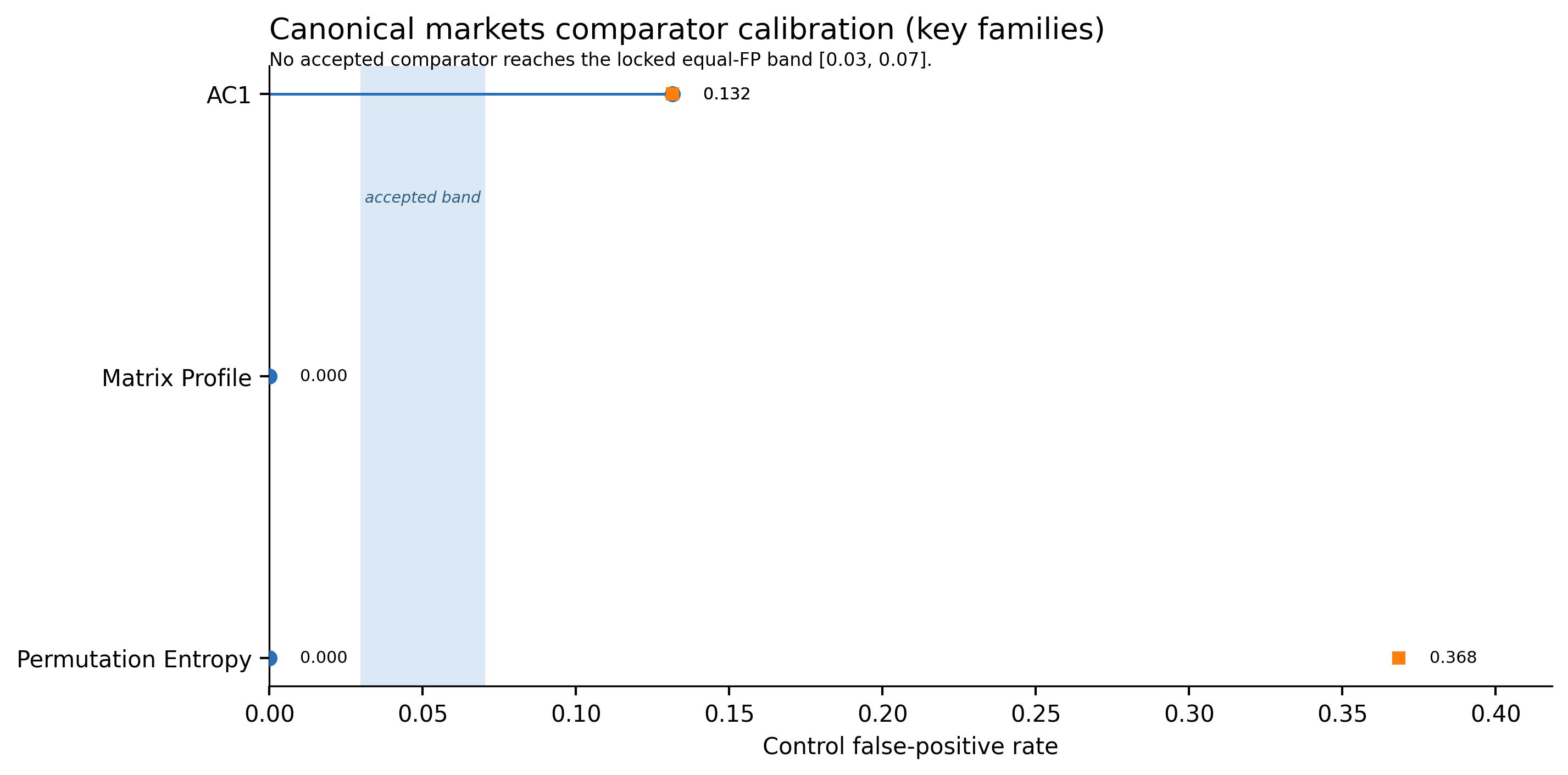}}

\textbf{Figure 4. Comparator calibration on the canonical segmented
markets benchmark (n=38 control units + 16 event units).} No tested fast
or slow comparator configuration reaches the locked equal-false-positive
band {[}0.03, 0.07{]}. The nearest nontrivial comparator is AC1 at FP =
0.131579 (5/38 control alarm units), while the numerically nearest slow
configurations are trivial-silent and the best nontrivial slow family,
permutation entropy, remains materially above band.

\subsubsection{Fast-family robustness under canonical and relaxed band
specifications}\label{fast-family-robustness-under-canonical-and-relaxed-band-specifications}

The comparator conclusion in markets admits a narrower and more precise
robustness statement. Under the prespecified equal-false-positive band
{[}0.03, 0.07{]}, no tested fast-family comparator configuration was
accepted on the canonical 120-minute segmented markets benchmark or on
60-minute and 180-minute segmentation sensitivity variants. In all three
cases, the nearest fast family remained AC1; the nearest false-positive
rate was 0.1316 for the canonical 120-minute units, 0.0800 for the
60-minute units, and 0.1304 for the 180-minute units. Thus, fast-family
non-acceptance was robust across tested segmentations at the locked
canonical specification.

By contrast, post-processing band-sensitivity analysis showed that
widening the acceptable interval to {[}0.02, 0.08{]} or {[}0.04, 0.08{]}
admitted the 60-minute AC1 configuration, whereas the canonical
120-minute and 180-minute segmentations remained non-accepted. We
therefore distinguish canonical robustness from relaxed-band sensitivity
and do not claim full invariance to post hoc band relaxation. The result
that carries manuscript weight is the stronger one: at the prespecified
canonical band, no fast-family comparator is accepted at 60-, 120-, or
180-minute segmentation.

\subsubsection{Controls, falsification boundary, and bounded
exceptions}\label{controls-falsification-boundary-and-bounded-exceptions}

The bridge criterion is explicitly falsifiable. It would be weakened in
four ways. First, if externally defined collapse events occurred without
prior increase in G and p and decrease in δ. Second, if comparator
configurations detected the same events under matched false-positive
constraints without exhibiting this triad. Third, if the (G, p, δ)
pattern appeared frequently in control periods without progression
toward externally defined collapse. Fourth, if alternative reasonable
operationalizations of amplification, recursive persistence, and
diversity contraction failed to agree directionally with the present
measurements. At the benchmark level, the comparator non-recovery claim
would fail if any tested comparator configuration achieved an accepted
operating point on the canonical benchmark under the same
equal-false-positive contract.

The markets branch also contains a bounded near-miss rather than a
universal separator. Binary and topological diagnostics did not separate
Volmageddon from vol-heavy hard negatives. The strongest remaining
signal instead appeared in continuous witness quality, with the clearest
confirmation in the corroborating intraday-grain witness configuration
\texttt{cfg\_002053}, particularly against the post-Volmageddon
volatility-cluster control segment of February 8, 2018
(\texttt{\seqsplit{volmageddon\_control\_2018\_02\_08}}). We therefore treat this
residual as a bounded exception that sharpens the empirical
interpretation rather than as a contradiction of the broader result.
Across the retained domains, neither the bridge criterion nor the
comparator non-recovery claim is falsified.

\subsubsection{Third-domain corroboration: LLM training-loop collapse
(secondary
analysis)}\label{third-domain-corroboration-llm-training-loop-collapse-secondary-analysis}

\pandocbounded{\includegraphics[keepaspectratio]{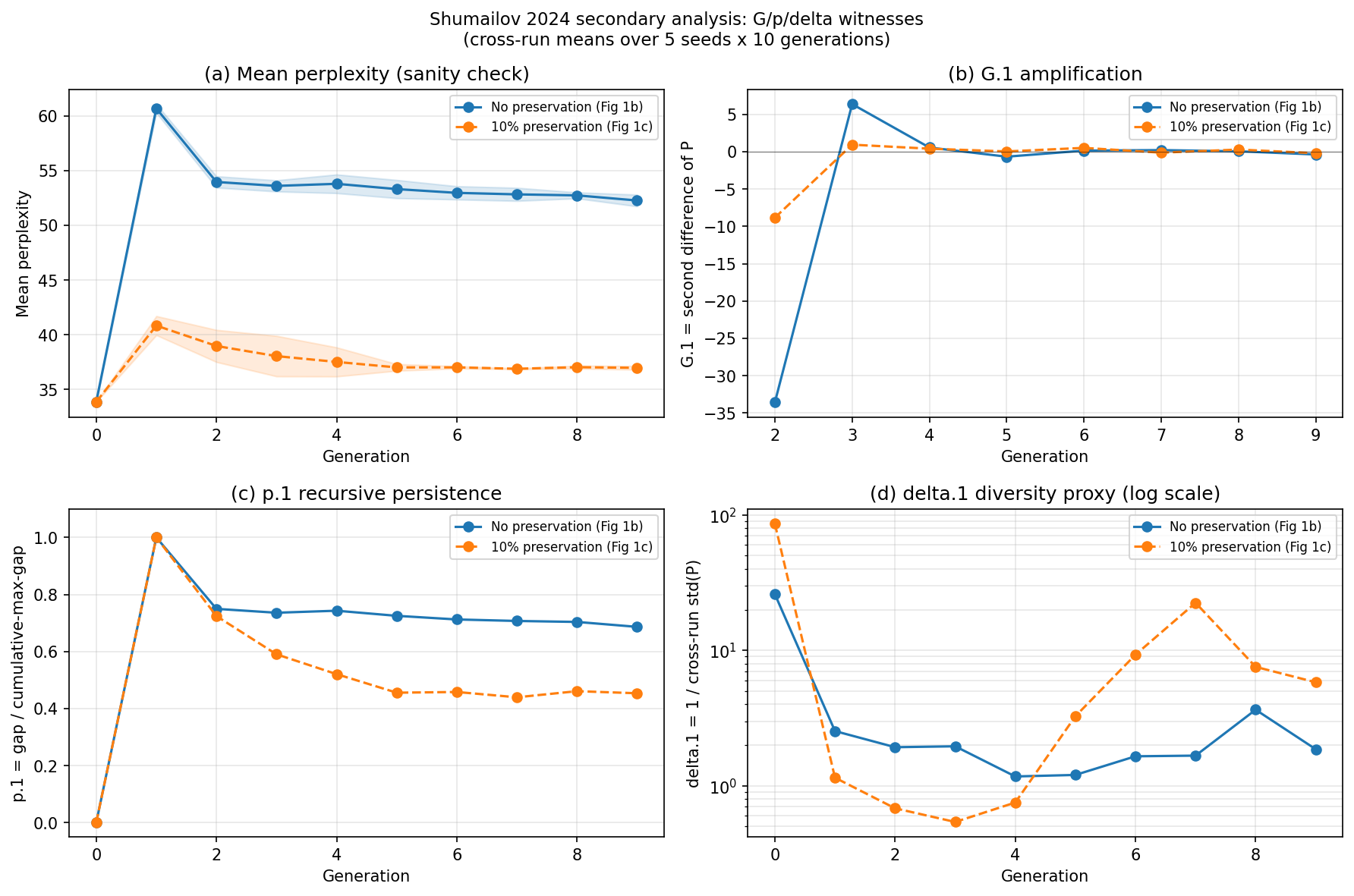}}

\textbf{Figure 5.} G/p/δ witnesses across 10 generations of recursive
LLM fine-tuning, secondary analysis on data digitized from Shumailov et
al.~(2024) Figure 1b/1c right panels. Lines show cross-run means over 5
random-seed runs. (a) Mean perplexity sanity check with ±1σ envelope.
(b) G.1 amplification (second difference of mean perplexity per run);
large negative deflection at generation 2 marks the phase-transition
signature. (c) p.1 recursive persistence (gap-to-cumulative-max ratio);
jumps to 1 at collapse onset and stays elevated. (d) δ.1 diversity proxy
(inverse cross-run perplexity standard deviation, log scale); declines
sharply at onset in both regimes. Two regimes shown: no preservation
(blue, Fig 1b) and 10\% real-data preservation (orange, Fig 1c).

Recursive language-model training exhibits the same closed-loop topology
as the markets and recommender domains: model outputs become inputs to
the next generation's training, creating conditions for
self-amplification and diversity contraction. Shumailov et al.~(2024)
demonstrated this empirically, showing that perplexity on held-out text
rises sharply within one generation of fine-tuning on synthetic data and
stabilizes at an elevated plateau, with magnitude depending on whether
real data is preserved across generations. We use this third recursive
domain as a directional-consistency check for the G/p/δ witness triad,
conducted as secondary analysis rather than as a matched-FP benchmark.

We digitized mean perplexity values from Figure 1b/1c right panels of
Shumailov et al.~(five random-seed runs across ten recursive
generations, in two preservation regimes: no preservation and 10\%
preservation). Each witness was computed from per-generation perplexity
data: G.1 as the second difference of mean perplexity per run, p.1 as
the ratio of the current perplexity-gap to the per-run
cumulative-maximum gap, and δ.1 as the inverse of cross-run perplexity
standard deviation at each generation. Generation 0 (the Real wikitext2
baseline) is the control window; generations 1 through 9 are the event
window. The Shumailov G.1 operationalization captures phase-transition
magnitude (\textbar G\textbar{} at onset) rather than the
signed-elevation form used in the markets and recommender domains,
reflecting that perplexity deviations in either direction can signal a
regime shift in this setting.

All three witnesses move in the predicted direction across the collapse
transition in both regimes (Figure 5). G.1 shows a sharp
phase-transition signature: \textbar G\textbar{} at generation 2 is
22.6× the steady-state mean in the no-preservation regime and 16.7× in
the 10\%-preservation regime. p.1 jumps from 0 at baseline to
approximately 1 at generation 1 and never relaxes below 0.36 in either
regime. δ.1 declines sharply at onset (control/event ratio 13.3× and
15.0×). The framework generates three graded predictions, all of which
the data also support: G.1 transition magnitude scales with collapse
severity (3.8× larger without preservation); p.1 plateau height encodes
the regime (0.75 versus 0.45); δ.1 partially recovers in the
10\%-preservation regime as random seeds re-converge to a stable
collapsed attractor.

We do not claim a third benchmark. Values are digitized from published
figures rather than provided by the original authors, the δ.1
operationalization captures inter-seed convergence rather than the
within-model distributional diversity that the paper's Theorem 3.1
directly addresses, and no comparator-acceptance evaluation under a
matched false-positive contract was conducted for this domain. The full
benchmark - including experimental pipeline reconstruction at scale and
matched-FP evaluation - is deferred to follow-up work.

\subsubsection{Cross-benchmark synthesis: effect sizes and
comparator-band
ROC}\label{cross-benchmark-synthesis-effect-sizes-and-comparator-band-roc}

Figure 6 summarizes effect-size magnitudes and confidence intervals
across all four benchmark/horizon configurations and all three
witnesses. The visualization makes three findings inspectable at a
glance. First, the canonical recommender horizon h=50 (blue, anchored by
background tint) is the only configuration where all three witnesses
align in the predicted direction with confidence intervals clear of the
null reference. Second, adjacent recommender horizons degrade
asymmetrically rather than uniformly: h=40 shows a wrong-direction
effect on G with predicted-direction effects on p and δ, while h=60
shows a strong predicted-direction effect on G but null intervals on p
and δ. Third, markets effects at the per-row grain are uniformly null
across all three witnesses; the directional bridge claim for markets
operates at the unit-level aggregation grain reported in the late-window
analysis. BCa and percentile confidence intervals agree to within Monte
Carlo noise on all recommender cells (n = 40,339); they diverge
meaningfully on markets cells, where the right-skewed bootstrap
distribution at low n favors the BCa correction as the more honest read.
Full effect-size numerics, including Glass's d and rank AUC alongside
Cohen's d, are reported in Supplementary Table S2.

\pandocbounded{\includegraphics[keepaspectratio]{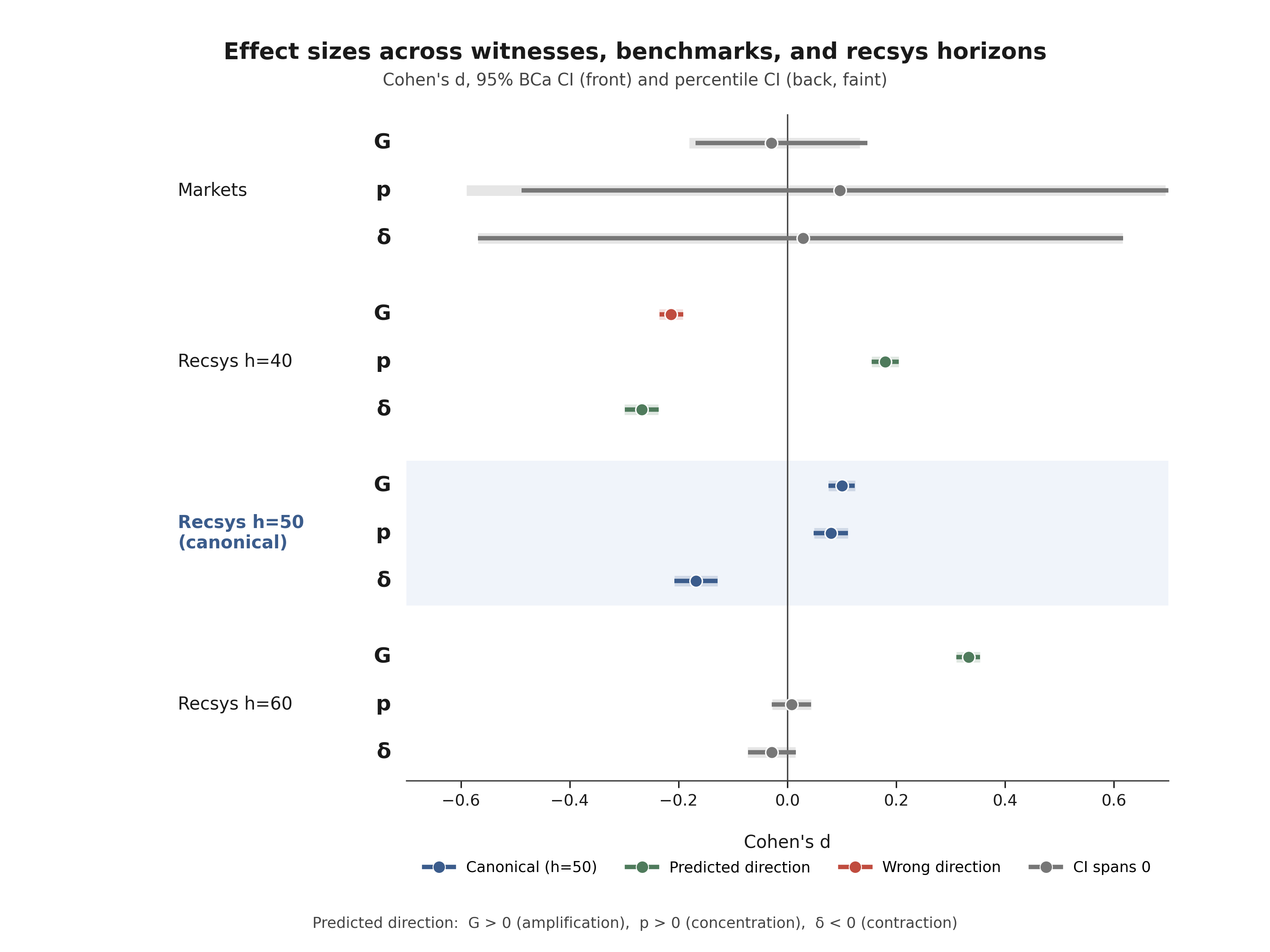}}

\textbf{Figure 6. Effect sizes across witnesses, benchmarks, and
recommender horizons.} Cohen's d point estimates with 95\%
bias-corrected accelerated (BCa) confidence intervals in the foreground
and 95\% percentile confidence intervals in the faint background.
Cluster-aware bootstrap (10,000 iterations) at segment-level grain for
markets (n=38 controls + 16 events) and user-level grain for recommender
(n = 40,339 user clusters per horizon panel; per-horizon event/control
splits --- h=40: n\_control=6,562 / n\_event=33,777; h=50 canonical:
n\_control=4,755 / n\_event=35,584; h=60: n\_control=3,996 /
n\_event=36,343; \textasciitilde10 rows per cluster). Color coding: blue
for the canonical h=50 group (anchored by background tint), green for
predicted-direction effects with CIs clear of null, red for
wrong-direction effects with CIs clear of null, gray for CIs spanning
null. Predicted direction: G \textgreater{} 0 (amplification), p
\textgreater{} 0 (concentration), δ \textless{} 0 (contraction). The
pre-registered canonical h=50 configuration is the only horizon where
all three witnesses align in the predicted direction with intervals
clear of null.

Figure 7 complements the effect-size synthesis with a direct
visualization of how comparator families occupy the false-positive axis
relative to the locked equal-FP band. Figure 6 quantifies how strongly
witnesses separate event from control telemetry; Figure 7 quantifies how
far each comparator family operates from the contract-mandated
acceptance region. Two findings sharpen the comparator-failure story.
First, on the late-30-minute ROC slice the markets slow families (matrix
profile, permutation entropy) fail at the data-availability gate before
reaching calibration: this short window leaves n\_control\_units = 6,
below the discrete-grid floor required to reach the {[}0.03, 0.07{]}
band, so they were filtered as \texttt{\seqsplit{unreachable\_fp\_band}} on this
slice. These families are evaluated on the full-window canonical markets
benchmark (Figure 4), where the nearest nontrivial slow configuration,
permutation entropy, reaches 14/38 control units at FP = 0.368, above
band. Second, the recommender slow families occupy the low-FP region but
at very low TPR --- visible in the inset zoom --- making the FP/TPR
trade-off inadequate for the contract at any horizon. The pre-registered
Loopzero conservative operating point (q=95, k=3) is marked as a gold
star at (FP ≈ 0, TPR ≈ 0); the achievable Loopzero envelope and its
matched-FP comparison against the comparator suite are reported in the
\emph{Envelope-boundary matched-FP comparison} subsection below.

\pandocbounded{\includegraphics[keepaspectratio]{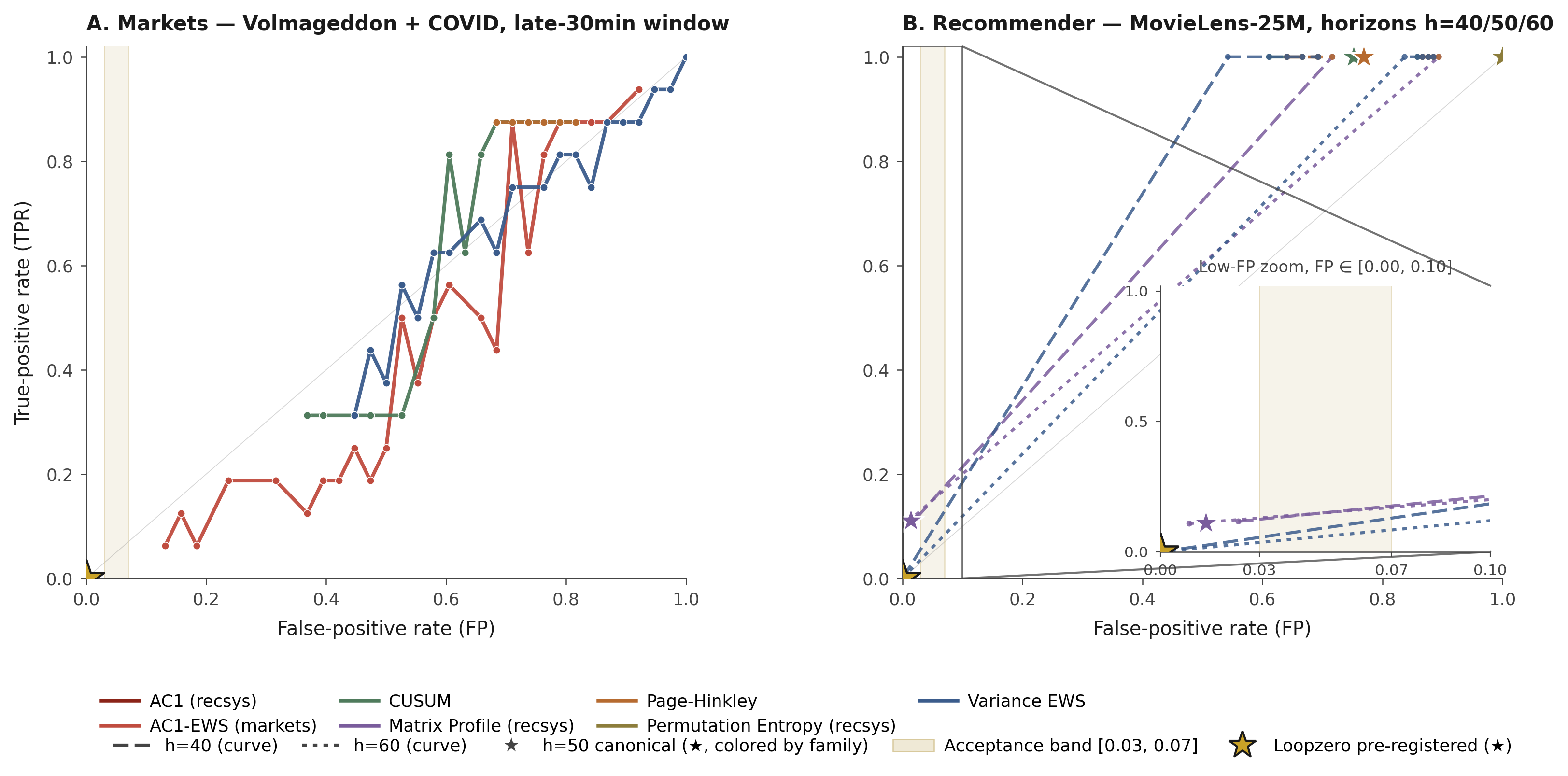}}

\textbf{Figure 7. Comparator operating points relative to the locked
equal-FP band {[}0.03, 0.07{]}.} \textbf{(A)} Markets (Volmageddon +
COVID, late-30min window; n=38 controls + 16 events): four fast families
overfire above the band; slow families (matrix profile, permutation
entropy) fail at the data-availability gate on this late-window panel
because their min-length requirement reduces usable n\_control\_units
below band reachability (these families are evaluated on the full-window
benchmark in Figure 4). \textbf{(B)} Recommender (MovieLens-25M; 40,339
user clusters per horizon panel; per-horizon event/control splits ---
h=40: n\_control=6,562 / n\_event=33,777; h=50 canonical:
n\_control=4,755 / n\_event=35,584; h=60: n\_control=3,996 /
n\_event=36,343) at horizons h=40 (long dash), h=50 (★ canonical), h=60
(short dash): all six families miss the band; inset shows the low-FP
region {[}0.0, 0.10{]} with the matrix-profile near-miss at FP ≈ 0.014.
Gold ★ marks the pre-registered Loopzero conservative operating point
(q=95, k=3) at (FP ≈ 0, TPR ≈ 0); this is a single deliberately strict
configuration, not the achievable Loopzero envelope. The achievable
Loopzero envelope across the extended envelope-boundary sensitivity grid
is reported in the \emph{Envelope-boundary matched-FP comparison}
subsection. No tested comparator configuration achieves an accepted
operating point under the locked equal-FP contract.

\paragraph{Envelope-boundary matched-FP
comparison}\label{envelope-boundary-matched-fp-comparison}

Figure 7 anchors the cross-benchmark synthesis at a single deliberately
strict Loopzero operating point (q = 95, k = 3) at (FP ≈ 0, TPR ≈ 0). To
characterize the achievable Loopzero envelope rather than this single
anchor, we extended the pre-registered sensitivity grid to q ∈ \{50, 60,
70, 75, 80, 85, 90, 95, 99\} (Methods), holding k = 3 fixed at the
canonical primary value. At each panel, Loopzero's envelope-boundary
operating point is the maximum control\_fp achieved across the extended
q-grid; comparator families are then reported at the linearly
interpolated TPR at the same boundary FP, or marked as a structural
non-overlap where the family's frozen calibration grid does not bracket
Loopzero's boundary (Table 2).

On the markets benchmark, Loopzero's envelope boundary at k = 3 is (FP =
0.237, TPR = 0.188), reached at q = 50. Of the four markets comparator
families that satisfied data-availability gating in the {[}0.03, 0.07{]}
equal-FP contract evaluation, only ac1\_ews has any operating point at
or below FP = 0.237; at that exact rate, the ac1\_ews and Loopzero
quantile detector operating points are identical. In integer alarm
counts (Supplementary Table S3), both methods alarm on exactly 9 of 38
controls and detect exactly 3 of 16 events at this matched anchor ---
alarm-set equivalence rather than a rate coincidence forced by the
discrete FP grid (control units quantize FP to multiples of 1/38 ≈
0.0263). The remaining three markets families (cusum, page\_hinkley,
variance\_ews) have minimum operating points strictly above Loopzero's
envelope-boundary FP --- they operate in a higher-FP regime than
Loopzero across the extended grid.

On the recommender benchmark, the pattern is sharper. Loopzero's
envelope boundary at k = 3 sits at FP = 0.0148 (h = 40), 0.0023 (h = 50
canonical), and 0.000501 (h = 60). Of the six standard EWS comparator
families calibrated on the recommender benchmark, four (ac1, cusum,
page\_hinkley, matrix\_profile) have minimum operating points that
exceed Loopzero's envelope boundary at h = 40 and h = 60; at h = 50,
Loopzero's boundary FP (0.0023) lies below the comparator calibration
grids' coverage, so h = 50 is excluded from this cross-horizon
comparison (Table 2, Supplementary Table S3). One family
(permutation\_entropy) has only a single unique FP per horizon in its
frozen calibration grid --- insufficient coverage to support envelope
comparison. Only one family (variance\_ews) has any envelope overlap
with Loopzero's quantile detector, and that overlap is itself qualified
by the comparator's pre-registered grid coverage: at h = 40,
variance\_ews's frozen calibration matrix has a 3559-alarm gap directly
across Loopzero's anchor of 97 alarms (Supplementary Table S3); at h =
60, the equivalent gap is 3345 alarms across Loopzero's anchor of 2
alarms. The linear-interpolated TPR values for variance\_ews at
Loopzero's envelope-boundary FP --- 0.027 at h = 40 and 0.0006 at h = 60
--- are accordingly estimates across these calibration-grid gaps rather
than measurements at the boundary FP.

The headline structural finding is therefore not
interpolation-dependent. Of the six standard EWS comparator families
calibrated on the recommender benchmark at the canonical k = 3, four
operate in a strictly higher false-positive regime than Loopzero's
quantile detector across all 9 q values of the extended
envelope-boundary sensitivity grid at h = 40 and h = 60. One has
insufficient calibration coverage to support envelope comparison. Only
variance\_ews has any calibration coverage in Loopzero's operating
regime --- coverage qualified by a 3559-alarm gap on h = 40 (3345 on h =
60) directly across Loopzero's envelope-boundary anchor. The
envelope-boundary contract therefore identifies a structural property of
standard EWS comparator literature on the recommender benchmark. The
operating regimes of the canonical comparator families do not extend
into the false-positive regions reached by a percentile-based
recursive-collapse detector across the extended envelope-boundary
sensitivity grid evaluated here.

\textbf{Table 2. Envelope-boundary matched-FP comparison at the
canonical k = 3.} Panels: markets (n=38 controls + 16 events) and
recommender at horizons h=40, h=50, h=60 (40,339 user clusters per
horizon panel; per-horizon event/control splits --- h=40:
n\_control=6,562 / n\_event=33,777; h=50 canonical: n\_control=4,755 /
n\_event=35,584; h=60: n\_control=3,996 / n\_event=36,343). Loopzero's
operating envelope boundary across the extended envelope-boundary
sensitivity grid (q ∈ \{50, 60, 70, 75, 80, 85, 90, 95, 99\}) on each
panel. Comparator families with envelope overlap are reported at the
linearly interpolated TPR at Loopzero's boundary FP; non-overlapping
families have minimum operating points exceeding Loopzero's boundary FP.
Alarm-count detail and breakpoint context in Supplementary Table S3.

{\def\LTcaptype{none} 
\begin{longtable}[]{@{}
  >{\raggedright\arraybackslash}p{(\linewidth - 6\tabcolsep) * \real{0.2500}}
  >{\raggedright\arraybackslash}p{(\linewidth - 6\tabcolsep) * \real{0.2500}}
  >{\raggedright\arraybackslash}p{(\linewidth - 6\tabcolsep) * \real{0.2500}}
  >{\raggedright\arraybackslash}p{(\linewidth - 6\tabcolsep) * \real{0.2500}}@{}}
\toprule\noalign{}
\begin{minipage}[b]{\linewidth}\raggedright
Panel
\end{minipage} & \begin{minipage}[b]{\linewidth}\raggedright
Loopzero envelope boundary at k=3
\end{minipage} & \begin{minipage}[b]{\linewidth}\raggedright
Overlapping comparator (TPR at boundary FP)
\end{minipage} & \begin{minipage}[b]{\linewidth}\raggedright
Non-overlapping families
\end{minipage} \\
\midrule\noalign{}
\endhead
\bottomrule\noalign{}
\endlastfoot
Markets & FP = 0.2368, TPR = 0.188 (q = 50) & ac1\_ews: 0.188 (exact
match; alarm-set equivalent at 9/38 controls, 3/16 events) & cusum,
page\_hinkley, variance\_ews: minimum FP \textgreater{} 0.237 \\
Recsys h=40 & FP = 0.0148, TPR = 0.092 (q = 50) & variance\_ews: 0.027
(interpolated across 3559-alarm calibration gap) & ac1, cusum,
page\_hinkley, matrix\_profile: minimum FP \textgreater{} 0.0148;
permutation\_entropy: insufficient grid coverage \\
Recsys h=50 & FP = 0.0023, TPR = 0.075 (q = 50) & --- (no comparator
breakpoint at this boundary FP) & --- \\
Recsys h=60 & FP = 0.000501, TPR = 0.055 (q = 50) & variance\_ews:
0.0006 (interpolated across 3345-alarm calibration gap) & ac1, cusum,
page\_hinkley, matrix\_profile: minimum FP \textgreater{} 0.0005;
permutation\_entropy: insufficient grid coverage \\
\end{longtable}
}

\subsection{Discussion}\label{discussion}

The scientific claim of the paper is deliberately minimal: collapse in
recursive systems admits a formally specified observable predicate with
an explicit claim boundary, and under equal-false-positive fairness,
tested standard comparator configurations did not recover the same
admissible operating behavior on canonical benchmarks. The significance
lies in the conjunction. A formal obstruction motivates a conditional
measurable pre-collapse signature; under matched-false-positive fairness
against a broad comparator suite, that directional signature is
recovered on the canonical benchmarks and no tested comparator
configuration achieved an accepted operating point. Unlike
threshold-tuned detectors, the empirical program is explicitly
falsifiable under a fixed equal-false-positive contract.

\emph{Methodological positioning.} The matched-false-positive contract
is conceptually rooted in operating-point selection on ROC curves and is
equivalent in spirit to Neyman--Pearson hypothesis testing evaluated at
a chosen specificity. In sequential change-point detection,
family-appropriate calibration of CUSUM and Page--Hinkley statistics is
conventionally done via expected average run length under the null
(\(\mathrm{ARL}_0\); Page, 1954); the locked equal-false-positive band
{[}0.03, 0.07{]} is the finite-sample matched-FP analog adapted for
recursive-collapse benchmarks where control-unit counts are bounded. The
contribution is not the invention of false-positive control but its
disciplined application: a prespecified band, frozen public-artifact
benchmarks with event labels fixed independently of witness computation
and comparator calibration, and reported outcomes including
non-acceptance.

\emph{Relation to early-warning-signal literature.} The framework is
distinct from the classical critical-transitions early-warning-signal
(EWS) framework. The canonical EWS suite --- increasing variance, lag-1
autocorrelation, and critical slowing down near bifurcations (Scheffer
et al., 2001, 2009) --- operates on stationary-noise dynamical systems
approaching a tipping point and has documented false-positive
vulnerabilities under non-stationary forcing (Boettiger \& Hastings,
2012a, 2012b). The matched-FP contract is the methodological response to
the family-comparability problem identified by that limitations
literature. The empirical bridge tested here is conceptually distinct:
G, p, and δ operationalize a recursive no-progress obstruction rather
than a noise-amplification signature near a bifurcation. Loopzero is
therefore not a replacement for critical-slowing-down detection in
dynamical systems near bifurcations; it tests a different conditional
bridge under an explicit alert-budget contract, on benchmarks where
recursive feedback structure is the primary feature.

\emph{Falsifiability and current verdict.} The framework is explicitly
falsifiable. It would be weakened if (i) externally defined collapse
events occurred without the predicted directional G/p/δ pattern in the
pre-collapse window; (ii) control units showed the same pattern without
subsequent collapse events; (iii) tested standard comparators recovered
accepted operating points under the locked false-positive contract on
the canonical benchmarks; or (iv) Loopzero's pre-registered quantile
detector recovered accepted operating points on benchmarks where the
bridge signature was absent. Across the canonical benchmarks, the
directional witness pattern is recovered in the pre-collapse windows; no
tested comparator configuration achieves an accepted operating point;
and Loopzero's pre-registered quantile detector itself does not achieve
an accepted operating point with nonzero event recovery on any panel.
This pattern is consistent with the framework as a falsifiable empirical
program rather than a completed operational detector.

\emph{Domain-specific restraint.} The recommender branch (MovieLens-25M
offline deterministic replay) provides a corroborating second flagship
benchmark. Offline recommender evaluation is subject to selection biases
relative to deployed-system behavior (Schnabel et al., 2016); results
test bridge survival under controlled deterministic replay rather than
predicting online user response. Adjacent-horizon sensitivity preserves
the comparator non-recovery result at h=40 and h=60 while the bridge
degrades asymmetrically (Supplementary Table S2), supporting a
corroborating benchmark with bounded sensitivity rather than unqualified
invariance.

The markets branch provides bounded real-world corroboration. The
segmented benchmark is externally grounded, the band is reachable, the
comparator suite is broad, and no tested fast or slow configuration
admits an accepted operating point. Inferential strength is limited by
small event-scenario count (n=2 event clusters) and within-event
dependence; cluster-aware bootstrap at scenario grain (Supplementary
Table S4) is reported as a dependence-sensitivity diagnostic, not as
primary inference. Wild cluster bootstrap (Cameron, Gelbach \& Miller,
2008) is the methodologically appropriate calibration in this
small-cluster regime and is deferred to follow-up work.

The third domain --- recursive LLM training-loop collapse (Shumailov et
al., 2024) --- was evaluated as a secondary directional-consistency
check rather than a matched-FP benchmark: all three witnesses moved in
the predicted direction across the collapse transition, with graded
predictions on collapse severity and preservation regime also visible.
Matched-FP comparator evaluation in this domain is deferred.

\emph{Limitations and forward path.} Witness construction is still
domain-adapted. Comparator scope could expand. The bridge criterion,
although explicit and falsifiable here, remains empirical rather than an
identification result. The threshold-path envelope-boundary analysis
shows that Loopzero's quantile detector operates in a structurally
lower-FP regime than four of the six standard recommender-benchmark
comparator families across the extended envelope-boundary sensitivity
grid at h = 40 and h = 60 --- a finding independent of detector tuning.
The directional bridge result and this envelope-boundary structural
finding remain distinct from the stronger binary-acceptance claim, which
the pre-registered (q=95, k=3) configuration does not satisfy and which
remains open for future detector development at parameter regimes
outside the present pre-registered grid. A per-witness effect-size
decomposition (Figure 6) indicates that no single witness carries the
canonical result. At the pre-registered horizon h=50, G, p, and δ each
align in the predicted direction with BCa intervals clear of null.
Diversity contributes the largest single magnitude there: Cohen's d ≈
−0.17, versus ≈ +0.10 for G and ≈ +0.08 for p, so diversity contraction
is the strongest individual witness on the canonical panel rather than a
passenger. The witnesses also trade off across horizons. At h=40, G runs
counter to prediction (d ≈ −0.21) while p and δ remain aligned with
intervals clear of null. At h=60, G strengthens (d ≈ +0.33) while the p
and δ intervals relax to null. On the markets benchmark the
segment-grain effect sizes are individually modest with intervals
spanning null, consistent with the localized late-window character of
the markets bridge. This decomposition describes relative per-witness
contribution and is not a detector ablation; component-level necessity
of G, p, and δ through full ablations is reserved for follow-up work.

The next scientific step is not rhetorical expansion but stress-testing:
additional domains under full matched-FP comparator evaluation, external
replication, intervention logic, and broader comparator classes.

\subsection{Materials and Methods}\label{materials-and-methods}

The paper combines a Lean-formalized minimal obstruction (de Moura \&
Ullrich, 2021; The mathlib Community, 2020) with empirical witness
construction over gain, recursive persistence, and diversity. Empirical
evaluation is performed under a locked equal-false-positive contract, so
detector families are compared at the same alert budget rather than at
arbitrary thresholds. Bridge satisfaction is operationally defined as
follows. Under the prespecified pre-collapse window, event-unit
summaries of G are higher than control-unit summaries, p maintains its
non-relaxation property, and event-unit summaries of δ are lower than
control-unit summaries. Bridge non-satisfaction is defined as failure of
any of these conditions. The remaining subsections specify the analysis
layers and claim status, domain inclusion criteria, per-domain data
provenance and benchmark construction, comparator scope, the
envelope-boundary matched-FP contract, cluster-robust sensitivity, the
BCa bootstrap procedure, artifact availability with axiom audit,
offline-evaluation limitations, and reproducibility anchors.

\textbf{Analysis layers and claim status.} The empirical and formal
contributions of this paper are organized into the following analysis
layers; each is reported under an explicit claim status to prevent
inference inflation across layers.

{\def\LTcaptype{none} 
\begin{longtable}[]{@{}
  >{\raggedright\arraybackslash}p{(\linewidth - 6\tabcolsep) * \real{0.2500}}
  >{\raggedright\arraybackslash}p{(\linewidth - 6\tabcolsep) * \real{0.2500}}
  >{\raggedright\arraybackslash}p{(\linewidth - 6\tabcolsep) * \real{0.2500}}
  >{\raggedright\arraybackslash}p{(\linewidth - 6\tabcolsep) * \real{0.2500}}@{}}
\toprule\noalign{}
\begin{minipage}[b]{\linewidth}\raggedright
Layer
\end{minipage} & \begin{minipage}[b]{\linewidth}\raggedright
Purpose
\end{minipage} & \begin{minipage}[b]{\linewidth}\raggedright
Pre-registered?
\end{minipage} & \begin{minipage}[b]{\linewidth}\raggedright
Claim status
\end{minipage} \\
\midrule\noalign{}
\endhead
\bottomrule\noalign{}
\endlastfoot
A0 & Lean-formalized recursive no-progress obstruction
(\texttt{\seqsplit{collapse\_via\_progresscycle\_public}},
\texttt{\seqsplit{telemetry\_bridge\_obstruction\_public}}) & Frozen Lean artifact
& Formal claim boundary only; does not verify empirical telemetry or
measurement map \\
A1 & Loopzero's primary quantile detector at canonical configuration (q
= 95, k = 3) under locked equal-FP band {[}0.03, 0.07{]} & Yes
(canonical configuration pre-declared) & Non-accepted on both benchmarks
(FP = 0.000, event alarm rate = 0.000) \\
A2 & Envelope-boundary threshold-path analysis over extended quantile
grid q ∈ \{50, 60, 70, 75, 80, 85, 90, 95, 99\} & Extended sensitivity
grid & Exploratory structural analysis; not a primary detector claim \\
A3 & G/p/δ witness Cohen's d effect sizes with cluster-aware BCa
intervals & Descriptive & Directional alignment and magnitude summary;
effect sizes small at h=50 canonical and weakening for h=60 on p and
δ \\
B1 & Negative recommender controls (shuffled-timestamp, popularity-only,
random, matrix-factorization, sequential) & Not yet run & Deferred to
v1.1; required for recursive-recommender mechanism specificity \\
B3 & Scenario-grain markets cluster sensitivity (n = 7 clusters;
Supplementary Table S4) & Sensitivity & Dependence-sensitivity
diagnostic at n = 2 event clusters; intervals not reliable conservative
inferential intervals; wild cluster bootstrap deferred \\
\end{longtable}
}

Acceptance under the locked false-positive contract is a property of A1
only. A2 results are reported to characterize the threshold-path
structure of the pre-registered detector grid and the comparator grids;
they are not used to claim operating-point acceptance. A3 effect-size
summaries support the directional bridge but are not equivalent to
detector acceptance.

\textbf{Domain inclusion criteria.} The matched false-positive framework
requires three operational criteria of a candidate domain. First, a
frozen public-artifact benchmark with event labels fixed independently
of the witness computation and comparator calibration --- either
externally defined real-world events (e.g., the markets-domain
intervention windows used here) or benchmark-defined held-out failures
under frozen rules (e.g., the MovieLens-25M held-out frontier failures
used here), in both cases not generated by the analytical framework
being tested. Second, sufficient control units to make the prespecified
equal-false-positive band {[}0.03, 0.07{]} reachable on the control-unit
FP grid. Third, feasible witness construction: a measurement map from
system state to a preorder under which the G, p, δ witness triad is
observable. Two recursive-feedback domains satisfy all three criteria
and are adopted as canonical matched-FP benchmarks: a markets domain
(volmageddon 2018 and COVID MWCB 2020) and a recommender-system domain
(MovieLens-25M offline deterministic replay). The recursive
language-model training-loop domain (Shumailov et al., 2024) is retained
as a secondary directional-consistency check only. Published
per-generation trajectories support directional witness analysis, but
raw trajectories at the scale required for matched-FP comparator
evaluation are not publicly available; matched-FP evaluation in this
domain is deferred to follow-up work. Other recursive-feedback domains
were considered but not adopted. Synthetic bandit simulators (toy
bandits, LinUCB, ε-greedy) were excluded because the collapse criterion
is part of the simulator specification, creating circularity under
matched-FP evaluation. A cognition domain (n-back and digit-span
working-memory tasks) was excluded because no public dataset at scale
provides externally validated working-memory-failure labels suitable for
matched-FP comparator evaluation. Cultural and social-media systems
(Wikipedia edit dynamics, Reddit thread evolution) were excluded because
candidate ``echo-chamber'' or ``topic-saturation'' labels co-occur with
the witness pattern by construction (topic-entropy floor functioning as
both label and δ-proxy), creating measurement circularity.

\textbf{Markets data provenance.} Minute-bar OHLCV data for the six
US-listed instruments SPY, QQQ, IWM, VXX, UVXY, SVXY across the
canonical 2017--2020 window --- covering both the XIV/SVXY February 2018
Volmageddon implosion and the March 2020 COVID-era Level 1 market-wide
circuit breaker triggers --- were obtained via the Alpaca Markets
historical market data API (\url{https://alpaca.markets/}, accessed
2025). Under the vendor's terms of service, raw minute bars are not
redistributed alongside this artifact; the ingestion and preprocessing
pipeline is maintained in an upstream private repository and reproduces
minute bars from an Alpaca subscription. Per-segment derived ingredient
packets containing log returns, absolute returns, intraday stress flags,
breadth indicators, and Loopzero telemetry features (G, p, δ) at
one-minute resolution are released in this artifact at
\texttt{\seqsplit{results/rendered/equity\_dislocation\_family/intraday\_v2\_ingredient\_packet/}};
each canonical event window (\texttt{\seqsplit{volmageddon\_2018\_xiv}},
\texttt{\seqsplit{covid\_mwcb\_2020\_03\_18}}) and matched control window is
materialized as a 54-column packet over the 16-hour intraday window
(typically 944 rows per segment). Event-window definitions, trigger
semantics, and the canonical late-30-minute summary slice follow the
post-event analyses cited in the references (Augustin et al., 2021; NYSE
Market-Wide Circuit Breaker Working Group, 2021). The matched-FP
comparator calibration runs against the frozen
\texttt{\seqsplit{markets\_comparator\_merged\_state\_v2}} (Supplementary Table
S5). The public-reproducibility scope of the markets benchmark is at the
derived-packet level: the released ingredient packets contain all
telemetry features (G, p, δ, breadth indicators, stress flags) and
event/control window labels required for any researcher to independently
re-run the matched-FP contract against the frozen comparator calibration
state. Bottom-up reconstruction from raw minute bars requires an Alpaca
Markets subscription under vendor terms; this scope partition ---
derived telemetry public, raw vendor feed private --- is standard for
vendor-licensed market data and does not affect verifiability of the
matched-FP comparator-acceptance results reported in this paper.

\textbf{Markets benchmark construction.} The canonical segmented markets
benchmark \texttt{\seqsplit{volmageddon\_covid\_public\_v2}} (canonical Loopzero
configuration \texttt{cfg\_001339}) was constructed from a curated set
of intraday packet ingredients drawn from minute-bar data for six
US-listed instruments --- SPY, QQQ, IWM, VXX, UVXY, and SVXY ---
covering the Volmageddon dislocation (February 2018; Augustin et al.,
2021) and the March 2020 COVID market-wide circuit-breaker cluster (NYSE
MWCB Working Group, 2021). Each canonical packet contains preprocessed
Loopzero telemetry (G, p, δ) for one underlying time series;
non-canonical configurations were excluded to avoid duplicated session
units. Each packet was sliced into 120-minute time-window segments
(minimum 60 retained minute-bars per segment), and each (slice ×
segment) pair became one comparator unit. Event/control assignment was
made by membership of the slice identifier in a prespecified list of
canonical event identifiers (e.g., \texttt{\seqsplit{volmageddon\_2018\_xiv}},
\texttt{\seqsplit{covid\_mwcb\_2020\_03\_18}}), with hard negative controls drawn
from non-event slices over the same instrument universe. Under this
construction, 38 control units and 16 event units were retained,
yielding an FP grid step of 1/38 = 0.026316. Calibration semantics are
full-benchmark with per-unit rolling-quantile thresholds and no held-out
partition. Because each segment is treated as an independent comparator
unit during calibration, intra-event correlation under shared
market-wide stress is a dependence limitation addressed at scenario
grain in the Cluster-robust sensitivity subsection (B3); wild cluster
bootstrap remains for follow-up work.

\textbf{Recommender benchmark construction.} The canonical recursive
recommender benchmark
\texttt{\seqsplit{movielens25m\_recursive\_frontier\_public\_v1}} was constructed
from the public MovieLens-25M dataset
(\url{https://grouplens.org/datasets/movielens/25m/}). Per-user
trajectories were sorted chronologically and reduced to one episode per
user. Each episode was replayed under a deterministic item-item
collaborative-filtering engine with a warm-start prefix and a held-out
positive frontier; collapse was defined as failure to recover held-out
positive-frontier items under frozen benchmark rules. The canonical
horizon was 50 recursive update steps; adjacent horizons of 40 and 60
steps were tested for bridge-sensitivity robustness. Under the canonical
50-step construction, 40,339 user-level units satisfied inclusion
criteria, of which 35,584 were event units (collapse occurred within the
canonical horizon) and 4,755 were control units (collapse did not occur
within the canonical horizon), yielding an FP grid step of 1/4755 ≈
0.0002103. Comparator families were evaluated against a derived per-user
benchmark series, \texttt{miss\_run\_fraction}, capturing recursive
frontier-miss runs over the canonical horizon. Each user is treated as
an independent unit in calibration. Full parameter specification ---
including the positive-rating threshold, frontier size, warm-start
prefix length, item-item engine configuration, and leakage-prevention
rules --- is recorded in the canonical manuscript freeze state
(\texttt{\seqsplit{movielens25m\_recursive\_frontier\_public\_v1\_\_manuscript\_freeze\_state.json}}),
with construction details documented in the Supplementary Materials.

\textbf{Comparator scope and exclusions.} The six comparator families
evaluated under the matched-FP contract --- variance EWS and lag-1
autocorrelation AC1 (Dakos et al., 2012), CUSUM (Page, 1954),
Page-Hinkley (Page, 1954), matrix profile (Yeh et al., 2016), and
permutation entropy (Bandt \& Pompe, 2002) --- are the established
detection-theoretic and early-warning-signal baselines in the
critical-transitions and online change-point literature for the
recursive-collapse setting. Four additional comparator families are out
of scope for this version with explicit rationale. (i) Generalized
likelihood ratio (GLR) tests require parametric distributional
assumptions on the pre-collapse generative process that the matched-FP
contract is designed to avoid, and are reserved for follow-up work with
non-parametric extensions. (ii) Bayesian online change-point detection
is computationally prohibitive at the canonical recsys grain (n = 40,339
user clusters × 50-step horizon); Rao-Blackwellized acceleration is
deferred to follow-up work. (iii) Supervised sequence classifiers
require labeled training data, violating the unsupervised early-warning
scope of the matched-FP framework. (iv) Transformer-based anomaly
detectors and modern foundation-model detectors are within scope
methodologically but require pre-training corpora whose collapse-content
composition is not auditable, creating a circularity risk that a
follow-up protocol can address via self-supervised representation
baselines decoupled from collapse labels. The matched-FP framework is
family-agnostic: any detector returning a per-unit alarm decision can be
evaluated under the same locked contract; researchers extending the
comparator suite should follow the pre-registered grid protocol and
frozen calibration freeze pattern at
\texttt{\seqsplit{results/frozen/comparators/}}.

\textbf{Envelope-boundary matched-FP contract.} Loopzero's
pre-registered sensitivity grid (q ∈ \{90, 95, 99\}; k ∈ \{1, 3, 5\})
was extended downward to q ∈ \{50, 60, 70, 75, 80, 85, 90, 95, 99\} for
envelope-boundary threshold-path evaluation. The extension stays within
the detector's percentile design space and stops at q = 50 --- the
semantic boundary of percentile-based threshold detection, below which a
trigger sits at or beneath the reference median. The canonical primary
configuration (q = 95, k = 3) is unchanged. Standard EWS comparator
families operate from their frozen pre-registered calibration matrices;
we do not extend these grids post-hoc, since doing so would either
compromise byte-exact reproducibility of the frozen state or constitute
research-grade extension outside the original method papers'
recommendations. At each panel and at the canonical k = 3, Loopzero is
reported at the maximum control\_fp it achieves across the extended
q-grid (its operating envelope boundary); each comparator family is
reported at the linearly interpolated TPR at the same boundary FP where
the family's calibration grid brackets that FP, or marked as a
structural non-overlap otherwise. A companion sensitivity analysis
re-expresses this comparison in integer alarm counts, surfacing where
discrete FP space forces rate coincidences into literal alarm-count
equivalences and where linear interpolation fills quantitative gaps in
the frozen pre-registered calibration grids. Full computation is at
\texttt{\seqsplit{analysis/20\_compute\_a2\_threshold\_path.py}} and
\texttt{\seqsplit{analysis/21\_compute\_a2\_alert\_count\_exact.py}}, with outputs
in \texttt{results/calibrated/}.

\textbf{Cluster-robust sensitivity at scenario grain.} The markets
benchmark's matched-control design induces a paired cluster structure (2
event scenarios totaling 16 segments; 5 control scenarios totaling 38
segments). To verify that the markets effect-size analysis is robust to
within-scenario dependence beyond the segment-level resampling reported
in Supplementary Table S2, we re-ran the cluster-aware bootstrap with
cluster grain shifted from segment-level (n = 54 segments) to
scenario-level (n = 7 clusters). Point estimates are identical across
cluster grains; only the bootstrap CIs change. At n = 2 event clusters
the standard cluster bootstrap is in the small-cluster regime where
Cameron, Gelbach, and Miller (2008) documented downward variance bias
relative to wild cluster bootstrap; we treat the scenario-grain
bootstrap as sensitivity rather than primary inference, with wild
cluster bootstrap reserved for follow-up work. Full cluster composition
and per-cell results are reported in Supplementary Table S4.

\textbf{BCa bootstrap procedure.} Bias-corrected and accelerated (BCa)
intervals follow Efron (1987). The bias-correction constant \(z_0\) is
computed as the inverse normal CDF of the proportion of bootstrap
replicates below the point estimate. The acceleration constant
\(\hat a\) is computed via jackknife resampling at the cluster grain
(segment-level for markets, user-level for recsys). Separately,
bootstrap replicates that are degenerate (e.g., zero variance from
same-scenario cluster resamples) are finite-filtered, and the per-cell
\texttt{n\_degenerate} count is recorded (Supplementary Table S4 reports
one such case at 2 of 10,000 iterations for the p\textbar Glass's d cell
at B3 scenario grain). Across all 27 recsys cells at n = 40,339,
\textbar BCa − percentile\textbar{} \textless{} 0.01 on every endpoint,
consistent with BCa's O(1/√n) second-order correction theory; the recsys
BCa report is therefore primarily for procedural consistency with
markets, where the right-skewed bootstrap distribution at finite n
favors BCa as the more honest read.

\textbf{Artifact availability.} The Lean keeper surface ---
\texttt{\seqsplit{no\_progress\_cycle\_public}},
\texttt{\seqsplit{no\_progress\_kcycle\_public}},
\texttt{\seqsplit{collapse\_via\_progresscycle\_public}}, \texttt{TelemetryState},
\texttt{telemetryμ}, \texttt{\seqsplit{telemetry\_bridge\_obstruction\_public}},
and \texttt{\seqsplit{no\_telemetry\_forbidden\_cycle\_public}} --- is publicly
archived at \url{https://github.com/davidmullett/loopzero-paper-public}
(commit \texttt{\seqsplit{7afd763f38182ca49b2016fe38eeefff4560cdf0}}) and is
reproducible under Lean v4.30.0-rc2 with Mathlib at commit
\texttt{\seqsplit{3ba1ec58ec69cd649b9e5c61485a98d1dd37a00f}}. The build runs
cleanly under \texttt{lake\ build}. An axiom audit verifies that the
three obstruction theorems (\texttt{\seqsplit{no\_progress\_cycle\_public}},
\texttt{\seqsplit{no\_progress\_kcycle\_public}},
\texttt{\seqsplit{collapse\_via\_progresscycle\_public}}) compile without any axiom
dependencies, while the two telemetry-bridge theorems
(\texttt{\seqsplit{telemetry\_bridge\_obstruction\_public}},
\texttt{\seqsplit{no\_telemetry\_forbidden\_cycle\_public}}) depend only on the
standard Mathlib trusted base (\texttt{propext},
\texttt{Classical.choice}, \texttt{Quot.sound}). No non-standard axioms
appear in the keeper surface.

\textbf{Offline-evaluation limitations.} The MovieLens-25M recommender
benchmark is subject to the standard offline-evaluation biases of
explicit-rating datasets. Ratings are missing-not-at-random: items
receive ratings conditional on a user having chosen to consume them, so
the observed-rating distribution is not representative of the user's
true preference distribution. Without logged exposure data,
item-exposure probabilities cannot be inferred (Schnabel et al., 2016;
Chaney et al., 2018), and popularity bias affects both the candidate set
and the held-out frontier evaluation (Fleder \& Hosanagar, 2009; Jiang
et al., 2019; Mansoury et al., 2020). The benchmark constructed here is
therefore explicitly framed as a deterministic offline replay over a
frozen public dataset, not as evidence about deployed recommender
systems exposed to live user feedback. Mechanism-specificity controls
--- random recommender, popularity-only, matrix-factorization,
sequential, and shuffled-timestamp baselines --- are deferred to v1.1
and are required for claims about recursive-recommender collapse
mechanism specifically, as distinct from the recursive-replay
frontier-failure pattern reported here.

\textbf{Reproducibility anchors.} All empirical results in this paper
are anchored to immutable frozen artifacts indexed by SHA-256 hash. The
recommender benchmark uses the
\texttt{\seqsplit{movielens\_recursive\_replay\_engine}} v1.0.0 (family
\texttt{\seqsplit{item\_item\_collaborative\_filtering}}, engine hash
\texttt{56c1cff225d60c09}) operating under contract SHA-256
\texttt{\seqsplit{2e256b255a7f074c1516d70315ebb216241a4a7e8aba2db88b194417705fd71d}};
all three tested horizons (h=40, h=50 canonical, h=60) are generated
parametrically from this engine and contract. The upstream MovieLens-25M
archive is the canonical GroupLens release (\texttt{ml-25m.zip}, archive
SHA-256
\texttt{\seqsplit{8b21cfb7eb1706b4ec0aac894368d90acf26ebdfb6aced3ebd4ad5bd1eb9c6aa}},
MD5-verified against the official GroupLens checksum at
\url{https://files.grouplens.org/datasets/movielens/ml-25m.zip.md5};
raw-input provenance manifest at
\texttt{\seqsplit{results/manifests/movielens25m\_recursive\_frontier\_public\_v1\_\_raw\_input\_provenance.json}}).
Frozen-state freeze-manifests for each empirical component are listed in
Supplementary Table S5; each frozen directory additionally contains a
\texttt{LOCK\_NOTE.md} documenting the scientific and editorial scope of
that freeze. The three recsys horizon panels (h=40, h=50 canonical,
h=60) are byte-exact deterministic derivations from the single benchmark
freeze state and contract under horizon parameter h ∈ \{40, 50, 60\};
given identical (freeze\_state, contract, h), the engine produces
identical outputs. Horizon-spanning coverage is anchored by the recsys
horizon sensitivity manifest (Supplementary Table S5).

\begin{center}\rule{0.5\linewidth}{0.5pt}\end{center}

\subsection{Supplementary Materials}\label{supplementary-materials}

Supplementary materials accompanying this article comprise:

\begin{itemize}
\tightlist
\item
  Supplementary Figure S1: Fast-family segmentation and band sensitivity
  on the canonical markets benchmark
\item
  Supplementary Table S2: Effect sizes with bootstrap 95\% confidence
  intervals
\item
  Supplementary Table S3: Alert-count exact-matching sensitivity check
\item
  Supplementary Table S4: Cluster-robust sensitivity at scenario grain
  (markets benchmark)
\item
  Supplementary Table S5: Reproducibility anchor hashes
\end{itemize}

These are submitted as an arXiv ancillary file and are also available at
the public repository
(\url{https://github.com/davidmullett/loopzero-paper-public}).

\subsection{Funding}\label{funding}

This research was conducted by the author as an independent investigator
and received no external funding.

\subsection{Author contributions}\label{author-contributions}

D.M. conceived the study, developed the theoretical framework,
implemented all software and analyses, and wrote the manuscript.

\subsection{Competing interests}\label{competing-interests}

The author is the founder of Loopzero, Inc., which has filed a U.S.
provisional patent application (No.~64/042,661, filed April 17, 2026)
covering monitoring and verification methods related to the subject
matter of this work; the author is the named inventor. The author
declares no other competing interests.

\subsection{Data and code
availability}\label{data-and-code-availability}

All public analysis code, derived telemetry packets, frozen calibration
outputs, and reproducibility anchors needed to reproduce the reported
analyses from the released artifacts are available at
\url{https://github.com/davidmullett/loopzero-paper-public}; raw vendor
market data and the upstream ingestion/preprocessing repository are not
redistributed (see Markets data provenance). The repository includes:

\begin{itemize}
\tightlist
\item
  All analysis scripts used to generate every figure and table
  (\texttt{analysis/})
\item
  Frozen calibration outputs and operating-point data
  (\texttt{results/})
\item
  Benchmark construction code for the markets and recommender domains
\item
  This manuscript's source and rendered outputs (\texttt{manuscript/})
\end{itemize}

External datasets used:

\begin{itemize}
\tightlist
\item
  MovieLens-25M is publicly available from GroupLens
  (\url{https://grouplens.org/datasets/movielens/})
\item
  The Shumailov et al.~(2024) data shown in Figure 5 was digitized from
  the published figures of the cited Nature paper
\end{itemize}

Reproducibility anchors documented in Materials and Methods include
Loopzero's pre-registered detector configuration (q\_G=95, q\_p=95,
q\_delta=5, k=3), the equal-false-positive contract band {[}0.03,
0.07{]}, the analysis seeds (RANDOM\_SEED=42 for A3 segment-grain
analyses, RANDOM\_SEED\_B3=43 for B3 scenario-grain analyses), and the
bootstrap iteration count (10,000 per cell).

\subsection{Figure and Table Legends}\label{figure-and-table-legends}

\textbf{Supplementary Figure S1. Fast-family segmentation and band
sensitivity on the canonical markets benchmark.} No fast-family
comparator was accepted at 60-, 120-, or 180-minute segmentation under
the prespecified equal-false-positive band {[}0.03, 0.07{]}. Under
widened post-processing bands with upper cutoff 0.08, acceptance
appeared only for the 60-minute AC1 configuration; the canonical
120-minute and 180-minute segmentations remained non-accepted across all
tested bands.

\textbf{Supplementary Table S2. Effect sizes with bootstrap 95\%
confidence intervals.} Per-witness Cohen's d, Glass's d, and rank AUC at
each benchmark/horizon (4 benchmarks × 3 witnesses × 3 effect measures =
36 cells), with both BCa-adjusted and percentile bootstrap confidence
intervals from cluster-aware resampling (10,000 iterations). Markets
uses segment-level resampling (n=38 controls + 16 events); recommender
(all horizons) uses user-level resampling (n = 40,339 user clusters).
Canonical h=50 group is bolded throughout. A sensitivity section flags
cells where BCa and percentile CI endpoints differ by ≥ 0.05 (markets
p-witness cells only --- the bootstrap distribution is right-skewed at
low n, and BCa is the more honest read there). Full numerics rendered
deterministically by
\texttt{\seqsplit{analysis/17\_render\_a3\_supplementary\_table\_s2.py}} from the
per-cell bootstrap output at
\texttt{\seqsplit{results/rendered/effect\_sizes/a3\_effect\_sizes\_full.csv}}.

\textbf{Supplementary Table S3. Alert-count exact-matching sensitivity
check.} Re-expression of the envelope-boundary matched-FP comparison
(main-text Table 2) in integer alarm counts. For each panel at the
canonical k = 3, Loopzero is reported at its envelope-boundary alarm
counts (n\_control\_alarmed, n\_event\_alarmed); each comparator
family's frozen pre-registered calibration grid is searched for the
operating point with the nearest n\_control\_alarmed value, with status
classified as \texttt{exact\_match} (comparator has an operating point
at Loopzero's exact alarm count), \texttt{bounded\_gap} (Loopzero's
alarm count falls between two adjacent comparator operating points, with
\texttt{\seqsplit{gap\_width\_n\_control}} quantifying the calibration coverage
gap), \texttt{no\_overlap\_above} (all comparator operating points have
higher alarm counts), or \texttt{insufficient\_data}. On the markets
benchmark, ac1\_ews exhibits \texttt{exact\_match} at Loopzero's
envelope-boundary anchor: both methods alarm on exactly 9 of 38 controls
and detect exactly 3 of 16 events. On the recommender benchmark,
variance\_ews exhibits the widest bounded gaps across Loopzero's
envelope boundary: 3559 alarms at h = 40 (Loopzero at 97 of 6562
controls) and 3345 alarms at h = 60 (Loopzero at 2 of 3996 controls), in
each case with the upper-bound comparator operating point detecting all
panel event units (TPR = 1.0). Full numerics rendered by
\texttt{\seqsplit{analysis/21\_compute\_a2\_alert\_count\_exact.py}} from frozen
calibration data at
\texttt{\seqsplit{results/rendered/a4\_roc\_lowfp/a4\_roc\_data.parquet}} and
operating-point data at
\texttt{results/rendered/bridge/a1\_loopzero\_operating\_points\{,\_h40\_h60\}.csv}.

\textbf{Supplementary Table S4. Cluster-robust sensitivity at scenario
grain (markets benchmark).} Per-witness effect sizes (Cohen's d, Glass's
d, rank AUC) on the markets benchmark at the A3 segment grain (n = 54
segments) and the B3 scenario grain (n = 7 clusters: 2 event scenarios
--- \texttt{\seqsplit{volmageddon\_2018\_xiv}} with 8 segments and
\texttt{\seqsplit{covid\_mwcb\_2020\_03\_18}} with 8 segments; 5 control scenarios
--- \texttt{\seqsplit{covid\_noncollapse\_2020\_03\_11}} with 8 segments,
\texttt{\seqsplit{volmageddon\_control\_2018\_01\_29}} with 8 segments,
\texttt{\seqsplit{volmageddon\_control\_2018\_02\_08}} with 8 segments,
\texttt{\seqsplit{covid\_noncollapse\_2020\_03\_13}} with 7 segments,
\texttt{\seqsplit{volmageddon\_control\_2018\_01\_25}} with 7 segments). Bootstrap
iterations: 10,000 per cell; CI level: 0.95; RNG seed for scenario
grain: 43 (distinct from A3 segment grain seed: 42). For
\texttt{p}\textbar Glass's d at scenario grain, 2 of 10,000 iterations
produced degenerate samples (control SD ≈ 0 from same-scenario cluster
resamples); finite-filter applied to remaining 9,998 iterations. All
other 8 cells: n\_degenerate = 0. Scenario-grain CIs are reported as
sensitivity, not primary inference; the segment-grain baseline
(Supplementary Table S2) remains the primary report.

\textbf{Supplementary Table S5. Reproducibility anchor hashes.}
Freeze-manifest SHA-256 hashes for the empirical components of this
paper, plus the recommender engine identifiers (engine\_hash
\texttt{56c1cff225d60c09}, contract SHA-256
\texttt{\seqsplit{2e256b255a7f074c1516d70315ebb216241a4a7e8aba2db88b194417705fd71d}})
and the upstream MovieLens-25M archive hash
(\texttt{\seqsplit{8b21cfb7eb1706b4ec0aac894368d90acf26ebdfb6aced3ebd4ad5bd1eb9c6aa}},
MD5-verified against the GroupLens official checksum). Full table is in
the Supplementary Materials.

\begin{center}\rule{0.5\linewidth}{0.5pt}\end{center}

\subsection{References}\label{references}

Augustin, P., Cheng, I.-H. and Van den Bergen, L. (2021). Volmageddon
and the failure of short volatility products. \emph{Financial Analysts
Journal}. \url{https://doi.org/10.1080/0015198X.2021.1913040}

Bandt, C. and Pompe, B. (2002). Permutation entropy: a natural
complexity measure for time series. Physical Review Letters, 88, 174102.
\url{https://doi.org/10.1103/PhysRevLett.88.174102}

Boettiger, C. and Hastings, A. (2012a). Early warning signals and the
prosecutor's fallacy. \emph{Proc. Royal Society B}.
\url{https://doi.org/10.1098/rspb.2012.2085}

Boettiger, C. and Hastings, A. (2012b). Quantifying limits to detection
of early warning for critical transitions. \emph{Journal of the Royal
Society Interface}. \url{https://doi.org/10.1098/rsif.2012.0125}

Brunnermeier, M.K. and Pedersen, L.H. (2009). Market liquidity and
funding liquidity. \emph{Review of Financial Studies}.
\url{https://doi.org/10.1093/rfs/hhn098}

Buldyrev, S.V., Parshani, R., Paul, G., Stanley, H.E. and Havlin, S.
(2010). Catastrophic cascade of failures in interdependent networks.
\emph{Nature}. \url{https://doi.org/10.1038/nature08932}

Cameron, A.C., Gelbach, J.B. and Miller, D.L. (2008). Bootstrap-based
improvements for inference with clustered errors. \emph{The Review of
Economics and Statistics}, 90(3), 414-427.
\url{https://doi.org/10.1162/rest.90.3.414}

Carpenter, S.R., Cole, J.J., Pace, M.L., Batt, R., Brock, W.A., Cline,
T. et al.~(2011). Early warnings of regime shifts: A whole-ecosystem
experiment. \emph{Science}.
\url{https://doi.org/10.1126/science.1203672}

CFTC and SEC Staff (2010). Findings Regarding the Market Events of May
6, 2010. Report, SEC / CFTC.
\url{https://www.sec.gov/news/studies/2010/marketevents-report.pdf}

Chaney, A.J.B., Stewart, B.M. and Engelhardt, B.E. (2018). How
algorithmic confounding in recommendation systems increases homogeneity
and decreases utility. \emph{ACM RecSys}.
\url{https://doi.org/10.1145/3240323.3240370}

Dakos, V., Carpenter, S.R., Brock, W.A., Ellison, A.M., Guttal, V.,
Ives, A.R. et al.~(2012). Methods for detecting early warnings of
critical transitions in time series illustrated using simulated
ecological data. \emph{PLOS ONE}.
\url{https://doi.org/10.1371/journal.pone.0041010}

de Moura, L. and Ullrich, S. (2021). The Lean 4 Theorem Prover and
Programming Language. \emph{CADE-28}.
\url{https://doi.org/10.1007/978-3-030-79876-5_37}

Efron, B. (1987). Better Bootstrap Confidence Intervals. Journal of the
American Statistical Association, 82(397), 171-185.
\url{https://doi.org/10.1080/01621459.1987.10478410}

ESRB Advisory Scientific Committee (2019). Can ETFs Contribute to
Systemic Risk? Report, ESRB.
\url{https://www.esrb.europa.eu/pub/pdf/asc/esrb.asc190617_9_canetfscontributesystemicrisk~983ea11870.en.pdf}

Fleder, D. and Hosanagar, K. (2009). Blockbuster culture's next rise or
fall: The impact of recommender systems on sales diversity.
\emph{Management Science}. \url{https://doi.org/10.1287/mnsc.1080.0974}

Gao, J., Barzel, B. and Barabasi, A.-L. (2016). Universal resilience
patterns in complex networks. \emph{Nature}.
\url{https://doi.org/10.1038/nature16948}

Hanley, J.A. and McNeil, B.J. (1982). The meaning and use of the area
under a receiver operating characteristic (ROC) curve. \emph{Radiology}.
\url{https://doi.org/10.1148/radiology.143.1.7063747}

Holling, C.S. (1973). Resilience and stability of ecological systems.
\emph{Annual Review of Ecology and Systematics}.
\url{https://doi.org/10.1146/annurev.es.04.110173.000245}

Jäger, G. and Füllsack, M. (2019). Systematically false positives in
early warning signal analysis. \emph{PLOS ONE}.
\url{https://doi.org/10.1371/journal.pone.0211072}

Jiang, R., Chiappa, S., Lattimore, T., Gyorgy, A. and Kohli, P. (2019).
Degenerate feedback loops in recommender systems. \emph{AIES/ACM}.
\url{https://doi.org/10.1145/3306618.3314288}

Mansoury, M., Abdollahpouri, H., Pechenizkiy, M., Mobasher, B. and
Burke, R. (2020). Feedback loop and bias amplification in recommender
systems. \emph{CIKM}. \url{https://doi.org/10.1145/3340531.3412152}

NYSE Market-Wide Circuit Breaker (MWCB) Working Group (2021). Report of
the Market-Wide Circuit Breaker (MWCB) Working Group. Report, NYSE /
SEC.
\url{https://www.nyse.com/publicdocs/nyse/markets/nyse/Report_of_the_Market-Wide_Circuit_Breaker_Working_Group.pdf}

Page, E.S. (1954). Continuous Inspection Schemes. Biometrika, 41,
100-115. \url{https://doi.org/10.1093/biomet/41.1-2.100}

Scheffer, M., Carpenter, S., Foley, J.A., Folke, C. and Walker, B.
(2001). Catastrophic shifts in ecosystems. \emph{Nature}.
\url{https://doi.org/10.1038/35098000}

Scheffer, M., Bascompte, J., Brock, W.A., Brovkin, V., Carpenter, S.R.,
Dakos, V. et al.~(2009). Early-warning signals for critical transitions.
\emph{Nature}. \url{https://doi.org/10.1038/nature08227}

Scheffer, M., Carpenter, S.R., Lenton, T.M., Bascompte, J., Brock, W.,
Dakos, V. et al.~(2012). Anticipating critical transitions.
\emph{Science}. \url{https://doi.org/10.1126/science.1225244}

Schnabel, T., Swaminathan, A., Singh, A., Chandak, N. and Joachims, T.
(2016). Recommendations as Treatments: Debiasing Learning and
Evaluation. ICML.
\url{https://proceedings.mlr.press/v48/schnabel16.html}

Shumailov, I., Shumaylov, Z., Zhao, Y., Papernot, N., Anderson, R. and
Gal, Y. (2024). AI models collapse when trained on recursively generated
data. \emph{Nature}. \url{https://doi.org/10.1038/s41586-024-07566-y}

Shumailov, I., Shumaylov, Z., Zhao, Y., Papernot, N., Anderson, R. and
Gal, Y. (2025). Author Correction: AI models collapse when trained on
recursively generated data. \emph{Nature}, 640, E6.
\url{https://doi.org/10.1038/s41586-025-08905-3}

The mathlib Community (2020). The Lean mathematical library. \emph{CPP
(ACM SIGPLAN)}. \url{https://doi.org/10.1145/3372885.3373824}

Walker, B., Holling, C.S., Carpenter, S.R. and Kinzig, A. (2004).
Resilience, adaptability and transformability in social-ecological
systems. \emph{Ecology and Society}.
\url{https://doi.org/10.5751/ES-00650-090205}

Whaley, R.E. (2009). Understanding the VIX. \emph{Journal of Portfolio
Management}. \url{https://doi.org/10.3905/jpm.2009.35.3.098}

Yeh, C.-C.M., Zhu, Y., Ulanova, L., Begum, N., Ding, Y., Dau, H.A.,
Silva, D.F., Mueen, A. and Keogh, E. (2016). Matrix Profile I: All Pairs
Similarity Joins for Time Series: A Unifying View that Includes Motifs,
Discords and Shapelets. IEEE ICDM, 1317-1322.
\url{https://doi.org/10.1109/ICDM.2016.0179}

\end{document}